\documentstyle[12pt]{article}
\include{epsf}
\epsfverbosetrue
\def\lo{\langle 0 |}
\def\ro{ | 0 \rangle }

\def\gmf{\gamma _{5}}

\def\la{\langle }
\def\ra{ \rangle }

\newcommand{\beq}{\begin{equation}}
\newcommand{\eeq}{\end{equation}}
\newcommand{\bea}{\begin{eqnarray}}
\newcommand{\eea}{\end{eqnarray}}

\begin{document}
                                        \begin{titlepage}
\begin{flushright}
hep-ph/9808469
\end{flushright}
\vskip1.8cm
\begin{center}
{\LARGE
 Domain Walls and Theta Dependence in QCD\\       
\vskip0.6cm
with an Effective Lagrangian Approach    
}         
\vskip1.5cm
 {\Large Todd Fugleberg ,}
{\Large Igor Halperin} 
and 
{\Large Ariel Zhitnitsky}
\vskip0.5cm
        Physics and Astronomy Department \\
        University of British Columbia \\
 6224 Agricultural Road, Vancouver, BC V6T 1Z1, Canada \\ 
        {\small e-mail:
fugle@physics.ubc.ca \\ 
higor@physics.ubc.ca \\
arz@physics.ubc.ca }
\vskip1.0cm
{\Large Abstract:\\}
\end{center}
\parbox[t]{\textwidth}{
We suggest an anomalous effective Lagrangian which 
reproduces the anomalous conformal and chiral Ward identities 
and topological charge quantization in QCD. It is shown that
the large  $N_c $ Di Vecchia-Veneziano-Witten 
effective chiral Lagrangian is locally 
recovered from our results, along with 
$ 1/N_c $ corrections, after integrating 
out the heavy ``glueball" fields. All dimensionful 
parameters in our scheme are fixed in terms of the quark and 
gluon condensates and quark masses.
We argue that for a certain range of parameters, metastable vacua 
appear which are separated from the true vacuum of lowest energy
by domain walls. The surface tension of the wall is estimated,
and the dynamics of the wall is discussed. The U(1) problem and 
the physics 
of the pseudo-goldstone bosons 
at different $ \theta $ angles are addressed within the effective 
Lagrangian approach. Implications for axion physics,
heavy ion collisions and the development of the early 
Universe during the QCD epoch are discussed.     
}

\vspace{1.0cm}

                                                \end{titlepage}

\section{Introduction}

The effective Lagrangian techniques have proved to be 
a powerful tool in quantum field theory.
 Generally, there exist two different definitions of an 
effective Lagrangian. 
One of them is the Wilsonian effective 
Lagrangian describing the low energy dynamics of the lightest particles
in the theory. In QCD, this is implemented by effective chiral
Lagrangians (ECL's) for the pseudoscalar mesons, which are 
essentially constrained by the global non-anomalous $ SU(N_f) \times 
SU(N_f) $ and (for large $ N_c $) anomalous $U(1) $ chiral symmetries.
Another type of effective Lagrangian (action) is defined as  
the Legendre transform of the generating functional
for connected Green functions. This 
object is relevant for addressing the vacuum properties of the 
theory in terms of vacuum expectation values (VEV's) of composite
operators, as they should minimize the effective action.
Such an approach is suitable for the study of the dependence 
of the QCD vacuum on external parameters, such as the light quark masses 
or the vacuum angle $ \theta $. The lowest dimensional
condensates $ \la \bar{\Psi} \Psi \ra , \la G^2 \ra,
\la G \tilde{G} \ra $, which are the most essential for 
the QCD vacuum structure,
are related to the anomalously and explicitly broken conformal 
and chiral symmetries of QCD. Thus, one can study the vacuum of QCD
with an effective Lagrangian realizing at the tree level 
anomalous conformal and chiral Ward identities of the theory.
The utility of such an approach to gauge theories has been recognized 
long ago for supersymmetric (SUSY) models, where anomalous effective
Lagrangians were found for both the pure gauge case \cite{VY}
and super-QCD (SQCD) \cite{TVY}.

The purpose of this paper is a detailed analysis of 
the anomalous effective Lagrangian for QCD 
with  $ N_f $ light flavors and $ N_c $ colors 
(more precisely, of its potential part) which was suggested recently 
by two of us, and briefly described in \cite{QCD}. 
It was obtained as a generalization of an anomalous effective 
Lagrangian for pure YM theory which was proposed earlier in 
\cite{1}. The constructions of \cite{1} and \cite{QCD} can be viewed
as non-supersymmetric counterparts of the  
Veneziano-Yankielowicz (VY) effective potential \cite{VY} 
for SUSY YM theory and the Taylor-Veneziano-Yankielowicz (TVY)
effective potential for SQCD, respectively. The results 
obtained in \cite{1,QCD} reveal some striking similarities between
the supersymmetric and non-supersymmetric effective potentials
and the physics that follows. Notably, the effective potentials
for QCD and gluodynamics are holomorphic functions of their fields,
analogously to the SUSY case\footnote{It should 
be noted that the status of holomorphy
in SUSY and ordinary QCD is different.
In supersymmetric theories, holomorphy is a consequence
of supersymmetry and, moreover, holomorphic 
combinations $ G^2 \pm i G \tilde{G} $ are determined 
by the structure of the anomaly supermultiplet. 
On the contrary, holomorphy in the ordinary
YM theory (in a special sense)
follows provided we make some plausible 
assumptions which are shown to be self-consistent 
{\it a posteriori}, see Sect.2.}. Moreover, they have both 
``dynamical" and ``topological" parts - a structure which is 
similar to that of the (amended \cite{KS}) VY effective potential 
\cite{VY}. As will be discussed in detail below, this ``topological"
part of the effective potential turns out to be crucial for 
the analysis of the physical $ \theta $ dependence in QCD.

The interest in such an effective Lagrangian 
for anomalously broken conformal and chiral $ U(1) $ symmetries
is several-fold.
First, it provides a generalization of the large $N_c $ 
Di Vecchia-Veneziano-Witten (VVW) ECL \cite{Wit2} 
(see also \cite{ECL})
for the case of arbitrary
$ N_c $ after integrating out the massive ``glueball" fields.
At this stage, no holomorphy is present in the resulting 
effective chiral potential. 
In this way we arrive at
a Wilsonian effective Lagrangian for the light degrees of 
freedom consistent with the Ward identities of the theory and a 
built-in quantization of the topological charge (see below). 
One may note that in principle such 
a Wilsonian effective Lagrangian satisfying
all general requirements on the theory could also be  
written down without constructing first a more 
complicated anomalous effective potential including also the 
``glueball" degrees of freedom. 
The resulting effective potential is found to contain a sine-Gordon
term whose large $ N_c $ expansion reproduces the VWW effective 
potential in the vicinity of the global minimum, along with
$ 1/N_c $ corrections. The presence of such term in the 
effective potential implies that the theory sustains the domain 
wall excitations. This observation may be important in the 
contexts of cosmology and heavy ion collisions. Furthermore,
our two-step approach to the derivation 
of the effective chiral Lagrangian has an additional merit in that
all dimensionful parameters in our scheme are fixed in terms of the 
gluon and quark condensates and quark masses. (The only entries
which are not fixed in our scheme are two dimensionless 
integer-valued parameters related to the vacuum structure and 
$ \theta $ dependence of the theory. As their values are still a 
subject of some controversy, see Sect. 3, in most of the paper
we will keep them as free parameters.)
The absence of free dimensionful
parameters helps to better understand the origin of the 
$ \eta' $ mass (the famous U(1) problem). In particular, 
it yields a new mass formula for the $ \eta' $ for finite $ N_c $ in 
terms 
of quark and gluon condensates in QCD (see Eq.(\ref{12}) below).  
Second, it allows one to address related questions of 
the phenomenology of pseudoscalar mesons, such as 
$ \pi^0 - \eta - \eta' $ or $ \eta -\eta' $ mixing with no further 
phenomenological input.
Third, such an effective Lagrangian allows one
to address the problem of $ \theta $ dependence in QCD. In contrast 
to the approach of Ref.\cite{Wit2} which deals 
from the very beginning with the light chiral degrees of freedom and 
explicitly incorporates the $ U_{A}(1) $ anomaly without restriction
of the topological charge to integer values, in our method both 
the $ U_{A}(1) $ anomaly and topological charge quantization are 
included in the effective Lagrangian framework. 
After the ``glueball" fields are integrated out, the topological
charge quantization still shows up in the limit $ V \rightarrow
\infty $ via the presence of certain 
cusps in the effective potential, which 
are not present in the large $ N_c $ ECL of 
Ref.\cite{Wit2}. Analogous ``glued" effective 
potentials containing cusp
singularities arise in supersymmetric $ N = 1 $ theories when 
quantization of the topological charge is imposed \cite{KS,KSS,KKS}. 
As will be discussed below, these modifications are not essential
for the local properties of the effective chiral potential
in the vicinity of the global minimum.
 In this case, the results of Ref.\cite{Wit2} are 
reproduced
along with calculable $ 1/N_c $ corrections. On the other 
hand, for large
values of $ \theta $ and/or $ \phi_i $ our results deviate from those 
of \cite{Wit2}. Last but not least, the problem of the $ \theta $ 
dependence in QCD is directly relevant for the construction of a 
realistic axion potential that would be compatible with the 
Ward identities of QCD. This is because an axion potential $ V(a) $ 
can be obtained, provided the functional form of the vacuum energy
in QCD
$ E_{vac}(\theta) $ is known, by the formal substitution $ \theta 
\rightarrow a(x)/f_{a} $. 

Our presentation is organized as follows. In Sect. 2 we describe 
the approach of Refs. \cite{1} and \cite{QCD} to the construction
of an anomalous effective Lagrangian for pure YM theory and QCD,
respectively. Sect. 3 discusses different proposals to find 
the dimensionless integer-valued parameters that enter our results.
A particular method suggested in \cite{3} to fix these numbers 
in the context of pure YM theory is presented in an appendix
in a form adopted for the case of full QCD. In Sect. 4 we show 
how the heavy ``glueball" degrees of freedom in our effective potential
can be integrated out, thus yielding an effective chiral potential
for the light degrees of freedom. A correspondence with the VWW ECL
\cite{Wit2} is established. In Sect. 5 we discuss the vacuum properties,
$ \theta $ dependence and domain wall solutions in the resulting 
effective theory. Sect. 6 is devoted to an analysis of the U(1) 
problem and properties of the pseudo-goldstone bosons at zero and
non-zero $ \theta $. Sect. 7 deals with the implications 
of our results for the properties of the axion and a possible 
study of the $ \theta $ dependence and new 
axion search experiment at RHIC. We also discuss the possibility 
of baryogenesis at the QCD scale, which seems suggestive in view of 
our results.

\section{Anomalous effective Lagrangian for QCD}

We start with recalling the construction \cite{1} 
of the anomalous effective potential for pure YM theory (gluodynamics).
It is defined as the Legendre transform of the generating 
functional for zero momentum correlation functions of the 
marginal operators $  G_{\mu \nu} 
\tilde{G}_{\mu \nu} $ and $ G_{\mu \nu} 
G_{\mu \nu} $ which are fixed by the conformal anomaly in 
terms of the gluon condensate \cite{NSVZ,2}. The effective potential
is a  function of effective zero momentum fields $ h , \bar{h} $
which describe the VEV's of the composite 
complex fields 
$ H ,\bar{H} $ :
\beq
\label{1}
\int dx \, h = \la \int dx \, H \ra \; \; , \; \; 
\int dx \, \bar{h} =  \la \int dx \, \bar{H} \ra \; , 
\eeq
where 
\beq
\label{2}
H  = \frac{1}{2} \left( \frac{ \beta(\alpha_s)}{4 \alpha_s}
G^2  + i \, \frac{1}{\xi_{YM}} \frac{ \alpha_s}{4 \pi} 
G \tilde{G} \right) \; , \; \bar{H}   
= \frac{1}{2} \left( \frac{ \beta(\alpha_s)}{4 \alpha_s}
G^2  - i \, \frac{1}{\xi_{YM}} \frac{ \alpha_s}{4 \pi} 
G \tilde{G} \right)  \; .
\eeq 
and $ \beta(\alpha_s)
= - b_{YM} \alpha_{s}^2/(2\pi)
+ O(\alpha_{s}^3) $ ,  $ b_{YM} = (11/3) N_c $ 
is the 
Gell-Mann - Low $ \beta $-function for YM theory\footnote{
In what follows we will work with the one-loop $ \beta $-function.
However, most of the discussion below can also be formulated with 
formally keeping the full $ \beta$-function.} 
, and $ \xi_{YM} $
is a generally unknown parameter which parametrizes the correlation
function of the topological density
\beq
\label{2a}
\lim_{ q \rightarrow 0} \; 
i \int dx \, e^{iqx} \lo T \left\{ \frac{\alpha_s}{8 \pi} 
G \tilde{G} (x)  \, 
\frac{\alpha_s}{8 \pi} G \tilde{G} (0) \right\}  \ro_{YM} =    
 \xi_{YM}^2  \, \la \frac{
\beta(\alpha_s)}{ 4 \alpha_s} G^2 \ra_{YM} \; . 
\eeq
The two-point function (\ref{2a}) and other 
zero momentum correlation functions of $ \alpha_{s} G^2 , \alpha_s
G \tilde{G} $ are defined  via the Wick type T-product by 
the nonperturbative part of the partition function
 $ \log (Z(\theta)/Z_{PT}) $ (
$ Z_{PT} $ stands for a perturbatively defined partition 
function which does not depend on $ \theta $), where 
\beq
\label{z}
Z (\theta ) =  Z_{PT} \exp \left\{ - i V E_{v}(\theta) \right\}
 = Z_{PT} \, \exp \left\{ - i V  \lo 
\frac{ \beta(\alpha_s)}{ 16 \alpha_s}
G^2 \ro_{\theta} \right\} \; 
\eeq 
by differentiation with respect to the bare coupling 
constant $ 1/g_{0}^{2} $ and $ \theta $. In Eq.(\ref{z}) we used 
the fact that the vacuum energy is defined relatively to its
value in perturbation theory by a nonperturbative part of the 
conformal anomaly, for any $ \theta $. When the vacuum expectation value 
(VEV) in Eq.(\ref{z}) is defined in this way, its dependence 
on $ 1/g_{0}^2 $ is fixed by the dimensional transmutation formula
\beq
\label{dt}
\la -  \frac{ b \alpha_s}{ 8  \pi}
G^2 \ra = const \, \left[ M_R \exp \left( - \frac{8 \pi^2}{
b g_{0}^2} \right) \right]^4 \; .
\eeq
Here $ M_{R} $ is the ultraviolet cut-off mass, and the one-loop
$ \beta $-function is used. It is important to stress that different 
regularization schemes generally lead to different values of the 
constant in Eq.(\ref{dt}) but, once specified, the VEV (\ref{dt})
determines {\it all} zero momentum correlation functions
of $ \beta(\alpha_{s})/(4 \alpha_s) G^2 $, with perturbative tails 
subtracted. At $ \theta = 0 $, Eq.(\ref{2a}) follows from Eq.(\ref{z})
using the general relation \cite{1}
\beq
\label{tt}
\la  \frac{ \beta(\alpha_s)}{ 4 \alpha_s}
G^2 \ra_{\theta} = \la  \frac{ \beta(\alpha_s)}{ 4 \alpha_s}
G^2 \ra_{0} \; f(\theta) \;  \; , \; \; 
f(\theta) \equiv 1 - 2 \xi^2 \, \theta^2 + O(\theta^4) \; .
\eeq
The strong assumption made in \cite{1} was that Eq.(\ref{2a}) is 
actually covariant in $ \theta $, i.e. remains valid for any (at least,
small) value of $ \theta $.  
The assumption of covariance in $ \theta $ is reproduced {\it a 
posteriori} from the effective Lagrangian, and is thus self-consistent.
We note that covariance of Eq.(\ref{2a}) in $ \theta $ follows 
automatically within the approach suggested in \cite{2}.
In fact, it is this conjectured covariance of Eq.(\ref{2a}) in 
$ \theta $ that underlies the holomorphic structure of the
resulting effective potential (see Eq.(\ref{3}) below). Thus, we 
are not able at the moment to {\it prove} holomorphy, but instead
argue that it is there based on (i) self-consistency of this 
proposal (see Sect.3), and (ii) the possibility of 
comparing our final
formulas with the known results (such as the large $N_c $ effective 
chiral Lagrangian and anomalous Ward identities in QCD, see Sect.4), 
and the experience from the known models. 

The advantage of 
using the combinations (\ref{2})
is in the holomorphic structure of 
zero momentum correlation functions of  
operators $ G^2 \,  ,\, G \tilde{G} $ written
in terms of the $ H, \bar{H} $ fields \cite{1,3}:
\bea
\label{2b}
\lim_{q 
\rightarrow 0 } \, i \int dx 
e^{iqx} \lo T \{ H(x) \; H(0) \} \ro &=& - 4 \la H \ra \; , 
\nonumber \\
\lim_{ q \rightarrow 
0} \,  i \int dx 
e^{iqx} \lo T \{ \bar{H}(x) \; \bar{H}(0) \} \ro &=& - 4 
\la \bar{H} \ra \; , \\ 
\lim_{ q \rightarrow 
0} \,  i \int dx 
e^{iqx} \lo T \{ \bar{H}(x) \; H(0) \} \ro &=&  0 \; . \nonumber 
\eea
In what follows Eqs.(\ref{2b}) will be sometimes referred to as the 
anomalous Ward identities (WI's)\footnote{It should be noted 
that 
the relations (\ref{2b}) are not the genuine 
anomalous WI's, as Eq.(\ref{2a})
used in (\ref{2b}) is not a Ward identity
in the proper sense, see the discussion above.}.  
It can be seen that the n-point zero momentum correlation
function of the operator $ H $ equals $ (-4)^{n-1} \la H \ra $.
Multi-point correlation functions of the operator $ \bar{H} $ 
are analogously expressed in terms of its vacuum expectation value 
$ \la \bar{H} \ra $. At the same time,
it is easy to check that the decoupling of the fields $ H $ 
and $ \bar{H} $ 
holds for arbitrary n-point functions 
of $ H $, $ \bar{H} $. This is the origin of holomorphy of an effective
Lagrangian for YM theory,
which codes information on all
anomalous WI's.

One should note that the 
right hand side of the last equation in (\ref{2b})
does contain perturbative contributions proportional to regular 
powers of $ \alpha_s $. However, they are irrelevant for our purposes,
as we 
are only interested in 
the decoupling of the fields $ H $ and $ \bar{H} $ at the level
of nonperturbative $ O(e^{- 1/\alpha_s}) $ effects.  
Holomorphy of an effective potential for YM theory has the same 
status. Thus, in contrast to the supersymmetric case where holomorphy 
is an exact property of the effective superpotential, in the 
present case it only refers to a ``nonperturbative" effective potential
which does not include perturbative effects to any finite order in 
$ \alpha_s $. We assume that perturbative and nonperturbative 
effects can be separated, at least in principle or/and by some
suitable convention \cite{2,1}. As a result, a perturbatively
defined partition function $ Z_{PT} $ bearing  
a non-holomorphic dependence on $ g_{0}^2 $ decouples in zero
momentum correlation functions.
On the other hand, the $ \theta $ dependence 
appears only in the nonperturbative part
of $ Z_{\theta} $. Thus, the {\it nonperturbative}
vacuum energy depends only on a single complex combination
$ \tau = 1/g_{0}^2 + i \theta b q /( 32 \pi^2 p) $ \cite{2,3}. Indeed,
arguments based on renormalizability, analogous to those used 
in Ref. \cite{NSVZ} (see also Sect. 3), imply the relation
\beq
\label{pq}
 \la - \frac{ b \alpha_s}{ 8 \pi} G^2 \ra_{\theta} = 
const \; Re \; M_{0}^4 \exp \left( - \frac{32 \pi^2}{b g_{0}^2 }
- i \frac{q}{p} \theta \right) \equiv const \; Re \; M_{0}^4 
\exp \left( - \frac{32 \pi^2}{b} \tau \right)
\; .  
\eeq
(This expression coincides with the result obtained in \cite{1}
directly
from the effective Lagrangian (\ref{3}).) This is exactly the 
origin of the relations (\ref{2b}) which are obtained by 
differentiation
of $ \log (Z_{\theta}/Z_{PT} ) $ with respect to the 
holomorphic sources 
$ \tau, \bar{ \tau} $. Thus, once an assumption of separation
of perturbative and nonperturbative contributions in Eq. (\ref{z}) 
is made, a new complex structure emerges due to the nonperturbative
origin of the $ \theta $ parameter which combines with another 
parameter $ 1/g_{0}^2 $ into the unique complex combination $ \tau $.   
 
The final answer for the 
improved effective potential $ W(h, \bar{h}) $ 
(here 'improved' refers
to the necessity of summation over the integers $n, k $ in 
Eq.(\ref{3}),
see below) reads \cite{1}   
\bea
\label{3}
e^{- i V W(h,\bar{h}) } &=& \sum_{n = - \infty}^{
 + \infty} \sum_{k=0}^{q-1} \exp \left\{ - \frac{i V}{4}
\left( h \, Log \, \frac{h}{C_{YM}} + 
\bar{h} \, Log \, \frac{ \bar{h}}{
\bar{C}_{YM}} \right) \right. \nonumber \\ 
&+& \left. i \pi V \left( k + \frac{q}{p} \,  
\frac{ \theta + 2 \pi n}{ 2 \pi} \right) \frac{h - \bar{h}}{
2 i} \right\} \;  ,
\eea
where    
the constants $ C_{YM}, \bar{C}_{YM} $ can be 
taken to be real and expressed in terms 
of the vacuum energy in YM theory at $ \theta = 0 $, 
$  C_{YM} =  \bar{C}_{YM} = - 2 e E_{v}^{(YM)}(0) 
= - 2 e \la - b_{YM} \alpha_s/(32 \pi) G^2 \ra $, and $ V $ 
is the 4-volume.
The integer numbers $ p $ and $q $ are relatively prime and 
related to the 
parameter $ \xi $
introduced in Eq.(\ref{2a}) by $ q/ p = 2 \xi $. 
Thus, we expect that the parameter $ \xi $ defined in 
Eqs.(\ref{2}),(\ref{2a}) is a rational number. This expectation 
is motivated by the fact that it turns out to be the case in all
existing proposals to fix the value of $ \xi $, to be discussed 
in the 
next section, and by experience with supersymmetric models. 
(In all likelihood, irrational values of $ \xi $ would produce
a non-differentiable $ \theta $ dependence for YM theory.)
On general grounds, it follows that $ p = O(N_c) \, , \, q = 
O(N_{c}^0)
$. 
The symbol $ Log $ in 
Eq.(\ref{3}) stands for the principal branch of the logarithm.
The effective potential (\ref{3}) produces an infinite series
of anomalous WI's. By construction, it is a periodic
function of the vacuum angle $ \theta $
\footnote{As was explained in \cite{1}, $ \theta $
should appear in the effective Lagrangian in the combination
$ \theta + 2 \pi n $, as this combination arises in the 
original YM partition function when the topological 
charge quantization is explicitly imposed. This prescription 
automatically ensures periodicity in $ \theta $ with period
$ 2 \pi $. However, with an additional 
integer $ k $ in Eq.(\ref{3}) and the above way of 
introducing $ \theta $ into the effective potential,
one more invariance under the shifts $ \theta \rightarrow 
\theta - 2 \pi l / q $ arises. As  will be discussed in Sect.5
in the context of QCD, the choice $ q \neq 1 $ is not in 
contradiction with the known results concerning the 
$ \theta $ dependence for small values of $ \theta $.}.
 The effective potential 
(\ref{3})  
is suitable for a study of the YM vacuum as described above. 
  
The double sum over the integers 
$ n , k $ in
Eq.(\ref{3}) appears as a resolution of an ambiguity
of the effective potential as defined from the anomalous WI's.
As was discussed in \cite{1}, this ambiguity is due to the fact
that any particular branch of the multi-valued function
$ h \log (h/c)^{p/q} $, corresponding to some fixed 
values of $ n , k $, satisfies the anomalous WI's.
However, without the summation over the integers $ n , k $ 
in Eq.(\ref{3}), the effective potential would be   
multi-valued and unbounded from below. 
An analogous problem arises with the 
original VY effective Lagrangian. It was cured 
by Kovner and Shifman in \cite{KS}
by a similar prescription of summation over all branches of the 
multi-valued VY superpotential. Moreover, the whole 
structure of Eq.(\ref{3}) is rather similar to that of the (amended)
VY effective potential. Namely, it contains
both the ``dynamical" and ``topological" parts (the 
first and the second terms in the exponent, respectively).
The ``dynamical" part of the effective potential (\ref{3}) 
is similar to the VY \cite{VY}
potential $
\sim S \log (S/ \Lambda)^{N_c} $ (here 
$ S $ is an anomaly superfield), 
while the ``topological" part is 
akin to the improvement \cite{KS} of the VY effective potential. 
Similarly to the supersymmetric case, the infinite sum over $ n $ 
reflects
the summation over all integer topological charges in the 
original YM theory. The difference of our case from 
that of supersymmetric YM theory is that 
an effective potential of the form $ (1/N) \phi \log (\phi/\Lambda)^N
$ , as in the SUSY case, implies a simpler form of the 
``topological" term $ \sim 2 \pi i n /N \, (\phi - \bar{\phi}) $
with only one ``topological number" $ n $ which specifies the 
particular branch of the multi-valued logarithm. In our case,
we allow for a more 
general situation when the parameter $ N $ is a rational number 
$ N = p/q $. In this case we have two integer 
valued ``topological numbers"
$ n $ and $ k $, specifying the branches of the logarithm and 
rational function, respectively.
Our choice is 
related to the fact that some proposals to fix the values of 
$ p, q $
suggest that $ q \neq 1 $, see Sect. 3. (As 
follows from Eq.(\ref{pq}), the values of 
$ p,q $ are fixed if the $ \theta $ dependence is known.)
 One may expect that 
the integers $ p $ and $ q $ are related to
a discrete symmetry surviving the anomaly,
which may not be directly visible in the original fundamental
Lagrangian.

It should be stressed that the improved effective potential
(\ref{3}) contains more information in comparison to that 
present in the anomalous Ward identities just due to the 
presence of the ``topological" part in Eq.(\ref{3}).
Without this term Eq.(\ref{3}) would merely be a kinematical
reformulation of the content of anomalous Ward identities for 
YM theory. The reason is that the latter refer,  
as usual, to the infinite volume (thermodynamic) limit 
of the theory, where only one state of a lowest energy (for $ 
\theta $ fixed) survives. This state corresponds to one 
particular branch of the multi-valued effective potential
in Eq.(\ref{3}). At the same time, the very fact of 
multi-valuedness
of the effective potential implies that there are other vacua 
which 
should all be taken into consideration when $ \theta $ is varied.
When summing over the integers $ n \, , \, k $, we keep track of 
all (including excited) vacua of the theory, and 
simultaneously solve the problems
of multi-valuedness and unboundedness from below of the ``one-branch
theory". The most attractive feature of the proposed structure
of the effective potential (\ref{3}) is that the same summation
over $ n , k $ reproduces the topological charge quantization
and $ 2 \pi $ periodicity in $ \theta $ of the original YM theory.

 We now proceed to the generalization of  
Eq.(\ref{3}) to 
the case of full QCD with $ N_f $ light flavors and $ N_c $ colors.
In the effective Lagrangian approach, the light matter fields are 
described by the unitary matrix $ U_{ij} $ corresponding to the
$ \gmf $ phases of the chiral condensate: $ \la \bar{\Psi}_{L}^{i} 
\Psi_{R}^{j} \ra 
=  - | \la \bar{\Psi}_{L} \Psi_{R} \ra | \, U_{ij} $ with 
\beq
\label{4}
U = \exp \left[ i \sqrt{2} \, \frac{\pi^{a} \lambda^{a} }{f_{\pi}}  + 
i \frac{ 2}{ \sqrt{N_{f}} } \frac{ \eta'}{ f_{\eta'}}  \right] 
\; \; , \; \; 
U U^{+} = 1 \; ,
\eeq
where $ \lambda^a  $ are the Gell-Mann matrices of $ SU(N_f) $, 
$ \pi^a $ 
is the pseudoscalar octet, and 
$ f_{\pi} = 133 \; MeV $. As is well known \cite{Wit2}, the 
effective potential for the $ U $ field (apart from the mass term) is   
uniquely determined by the chiral anomaly, and 
amounts to the substitution 
\beq
\label{5} 
\theta \rightarrow \theta - i \, Tr \, \log U \; 
\eeq
in the topological density term in the QCD Lagrangian. The 
rule (\ref{5}) is valid for any $ N_c $. 
Note that for spatially independent vacuum fields $ U $ Eq.(\ref{5})
results in the shift of $ \theta $ by a constant.
This fact will be used below. Furthermore, in the sense 
of anomalous conformal Ward identities \cite{NSVZ} QCD reduces to 
pure YM theory when the quarks are ``turned off" with the 
simultaneous substitution $ \la G^2 \ra_{QCD} \rightarrow 
\la G^2 \ra_{YM} $ 
and $ b \equiv b_{QCD} \rightarrow b_{YM} $.
Analogously,  
an effective Lagrangian for QCD should transform to
that of pure YM theory when the chiral fields $ U $ are ``frozen".    
Its form is thus suggested by the above arguments and 
Eqs.(\ref{3}),(\ref{5}):
\bea
\label{6}
e^{- i V W(h, U) } = \sum_{n = - \infty}^{
 + \infty} \sum_{k=0}^{q-1} \exp \left\{ - \frac{i V}{4}
\left( h \, Log \, \frac{h}{2 e E} + 
\bar{h} \, Log \, \frac{ \bar{h}}{
2 e E } \right) \right. \nonumber \\ 
+ \left. i \pi V \left( k + \frac{q}{p} \,  
\frac{ \theta - i \log Det \, U + 2 
\pi n}{ 2 \pi} \right) \frac{h - \bar{h}}{
2 i} + \frac{i}{2}  V Tr( M U + h.c.) \right\} \;  ,
\eea
where $ M = diag (m_{i}  | \la \bar{\Psi}^{i} \Psi^{i} 
\ra | )$ 
and
the complex fields $ h , \bar{h} $ are defined as in Eq.(\ref{2})
with the substitution $ b_{YM} \rightarrow b = (11/3) N_c - (2/3)
 N_f $.
The integers $ p, q $ and parameter $ \xi $  
($ q/p = 2 \xi $) in (\ref{6}) 
are in general different from those standing
in Eq.(\ref{3}). Possible approaches to fix the values of parameters
$ p $ and $ q $ in gluodynamics and QCD will be discussed in the next 
section. 
 The constant $ E $ can be related to the gluon condensate in 
QCD: $ E = \la b \alpha_s /(32 \pi) G^2 \ra $, as will be clear
below. We note that 
the ``dynamical" part of the anomalous effective potential (\ref{6})
can be written as $ W_d + W_{d}^{+} $ where 
\beq
\label{7}
W_{d} (h, U)  = \frac{1}{4} \frac{q}{p} h \, Log \left[ \left( 
\frac{h}{2 e E} \right)^{p/q}
\frac{ Det \, U }{ e^{-i\theta} } \right] -  \frac{1}{2}
Tr \, M U \; , 
\eeq
which is quite similar to  
the effective potential \cite{TVY} for SQCD\footnote{ For an early 
attempt 
to search for holomorphy in the effective Lagrangian framework for 
QCD, see \cite{Gomm}. The problem with the approach of Ref.\cite{Gomm}  
was that the resulting  
effective potential 
was multi-valued and unbounded from below.
The prescription of summation over all branches of the multi-valued
action in Eqs. (\ref{3}), (\ref{6}) cures both problems.}. 
 
Let us now check that
the anomalous WI's in QCD are reproduced from Eq.(\ref{6}).
The anomalous chiral WI's are automatically satisfied with the 
substitution (\ref{5}) for any $ N_c $, in accord with \cite{Wit2}.
Further, it can be seen that the anomalous conformal WI's of 
\cite{NSVZ} for zero momentum correlation functions of operator $ G^2 $
in the chiral limit $ m_q \rightarrow 0 $ 
are also satisfied with the above choice of constant $ E $. 
This is obvious from Eq.(\ref{7a}), see below. As 
another important example, we 
calculate the topological susceptibility 
in QCD near the chiral limit from
Eq.(\ref{6}). For simplicity, we consider the limit of 
$ SU(N_f) $ 
isospin symmetry with $ N_f $ light quarks, $ m_{i} \ll 
\Lambda_{QCD} $.
For the vacuum energy for small $ \theta < \pi/q $ we obtain (see
Eq.(\ref{18}) below)
\beq
\label{7a}
 E_{vac} (\theta) = -E  + m \la \bar{ \Psi} \Psi \ra  N_{f}
\cos \left( \frac{\theta}{N_{f}} \right) + O(m_{q}^2)  \; . 
\eeq
Differentiating this expression twice with
respect to $ \theta $, we reproduce
the result of \cite{SVZ}:
\beq
\label{7b}
\lim_{ q \rightarrow 0} \; 
i \int dx \, e^{iqx} \lo T \left\{ \frac{\alpha_s}{8 \pi} 
G \tilde{G} (x)  \, 
\frac{\alpha_s}{8 \pi} G \tilde{G} (0) \right\}  \ro =  
- \frac{ \partial^{2} E_{vac}(\theta)}{ \partial \, \theta^{2}} = 
 \frac{1}{N_f} m \la \bar{ \Psi} \Psi \ra  + O(m_{q}^2) \; .
\eeq
Other known anomalous WI's of QCD can be reproduced from Eq.(\ref{6})
in a similar way. Therefore, we see that  
Eq.(\ref{6}) 
reproduces the anomalous conformal
and chiral Ward identities of QCD and gives the correct
$ \theta $ dependence for small values of $ \theta $, and in this 
sense passes the  
test for it to be the effective anomalous potential for QCD.
Further arguments in favor of correctness of Eq.(\ref{6})
will be given in Sect. 4, where we show  
that
Eq.(\ref{6}) correctly reproduces the VVW ECL \cite{Wit2}
in the vicinity of the global minimum
in the large $ N_c $ limit after integrating out the heavy
``glueball" degrees of freedom, and in addition yields an
infinite series of $ 1/N_c $ corrections. 
On the other hand, we will explain why we obtain a different 
behavior of the effective chiral potential for large 
values of the chiral condensate phases $ \phi_{i} $.  

\section{What are the values of parameters $ p $ and $ q $ ?}

In the previous section we have considered the anomalous 
effective potentials for YM theory and QCD, which involve
some integer numbers $ p $ and $ q $, with $ 2 \xi = q/p $, 
which were not specified so far. 
The purpose of this section is to describe different 
proposals
to fix the numbers $ p ,q $, which exist in the literature, and 
to make some comments on them. A related discussion can be found in 
the Appendix.

Historically, the first suggestion to fix the proper
holomorphic combinations of the fields $ G^2 $ and $ G \tilde{G} $
was formulated in Ref.\cite{NSVZ}. For the case of pure YM theory, 
the authors proposed that fields of definite dualities dominate
the vacuum, and therefore the correct holomorphic combinations
are $ G^2 \pm i G \tilde{G} $. 
Leaving aside the issue of justification
of this hypothesis, it is of interest to discuss what values of 
the parameters $ p $ and $q $ are implied in this scenario. As 
was argued in \cite{NSVZ}, 
the $ \theta $ dependence of the vacuum energy is fixed in 
this case
by the renormalization group arguments, since for the VEV's of 
interest
the net effect of the $ \theta $ term reduces to the redefinition
of the coupling constant
\beq
\label{cc}
\frac{1}{g_{0}^2} \rightarrow \frac{1}{g_{0}^2} + 
\frac{ i \theta }{ 8 \pi^2} \; , 
\eeq
which yields for the vacuum energy for small values of $ \theta $
\beq
\label{sd}
\la - \frac{ b \alpha_s}{ 8 \pi} G^2 \ra_{\theta} = 
const \; Re \; M_{0}^4 \exp \left( - \frac{32 \pi^2}{b g_{0}^2 }
+ i \frac{4 \theta}{b} \right) \; ,  
\eeq
which corresponds to $ q/p = 4/b $, see Eq. (\ref{pq}). 
 Thus, we see 
that the self-duality 
hypothesis of Ref.\cite{NSVZ} implies the values $ p = 11 N_c \, , 
\, q = 12 $, for generic odd values of $ N_c 
\neq 3 k $ with some integer $ k $. As the value of 
$ p $ determines the number of different 
non-degenerate vacua in the theory \cite{1}, we end up with $ 11 N_c $ 
vacua, which may look strange. This is in contrast to the case 
of supersymmetric YM theory where the holomorphic combinations 
are {\it known} to be  $ G^2 \pm i G \tilde{G} $, but the number of 
vacua is $ N_c $ for any number of colors. This may be understood
in terms of the renormalization group arguments similar to 
Eqs. (\ref{cc}),(\ref{sd}) (with the substitution $ \la G^2 \ra 
\rightarrow \la \lambda \lambda \ra $) as a   
result of the interplay between the integer
valued $ \beta $-function $ b_{SYM} = 3 N_c $, which is determined 
by the zero modes alone and has a geometrical meaning, and the 
dimension $ d = 3 $ of the gluino condensate. It appears that 
this conspiracy is very specific to supersymmetric theories. 
It is interesting to note in this reference that 
if for some reason only the zero mode contribution $ b = 4 N_c $,
instead of the full $ b = (4-1/3)N_c $,  
were to be retained in the $ \beta $-function, Eq.(\ref{sd}) would 
imply $ N_c $ vacua. However, we are unable at the moment to 
see any compelling reason why such substitution should be made.
Thus, it remains unclear whether or not the appealing choice $ 
G^2 \pm i G \tilde{G} $ , $ p = N_c $ can be compatible with the 
renormalization group and conformal anomaly for non-supersymmetric 
YM theory.

Another approach to the problem of the number of vacua and 
proper holomorphic combinations of the fields $ G^2 $ and 
$ G \tilde{G}
$ is based on the analysis of softly broken SUSY theories \cite{MV}, 
which is under theoretical control as long as the gluino mass is much 
smaller than the dynamical mass scale: $ m_{g} \ll \Lambda_{SYM} $. 
A rather detailed discussion of this scenario has been recently given
by Shifman \cite{Shifman} using supersymmetric gluodynamics 
with $ N_c = 3 $ as an example. In the limit of small $ m_g $ the 
VEV
of the holomorphic combination $  G^2 + i G \tilde{G} $ is 
proportional to the VEV  $  m_g \la \lambda \lambda \ra $ where the 
gluino condensate $ \la \lambda \lambda \ra $ is to be calculated 
in the supersymmetric limit $ m_{g} = 0 $. The $ \theta $ 
dependence of 
the latter is known \cite{SV}: $ \la \lambda \lambda \ra \sim 
\exp (i 
\theta/ N_c + 2 \pi k /N_c ) \, , k = 0,1, \ldots, N_c -1 $, which 
corresponds to $ N_c $ degenerate vacua. When $ m_g \neq 0 $, 
the vacuum degeneracy is lifted. For $N_c = 3 $ and $ \theta = 0 $, 
we have one state with negative energy $ E = - m_g \Lambda_{SYM}^3
$, and two degenerate states with positive energy $ E = (1/2)    
m_g \Lambda_{SYM}^3 $. The former is the true vacuum state of  
softly broken SUSY gluodynamics, while the latter are metastable 
states with broken $ CP$. The lifetime of the metastable states 
is large for small $ m_g $, and decreases as $ m_g $ approaches
$ \Lambda_{SYM}$ . When $ \theta $ is varied, the three states 
intertwine,
thus restoring the physical $ 2 \pi $ periodicity in $ \theta $.  
This picture suggests the values $ p = N_c $, $ q = 1 $.
 
The problem with the above SUSY-motivated scenario is that the 
genuine
case of pure YM theory corresponds to the limit $ m_g \gg 
\Lambda_{SYM} $ which is not controlled in this 
approach. Moreover, the 
conformal anomaly in softly broken SUSY gluodynamics
is different from that of pure YM theory. On the other hand, it is 
clear from the above discussion 
that the conformal anomaly and dimensional 
transmutation are very essential for the analysis of the 
$ \theta $ dependence in gluodynamics. Perhaps, it is worthwhile 
to mention that
the value $ p = 11 N_c $ follows also within a non-standard 
non-soft SUSY breaking suggested recently \cite{SS} as a toy model
to match the conformal anomaly of non-supersymmetric YM theory
at the effective Lagrangian level. 

With these reservations, it is nevertheless reasonable to expect 
that the above SUSY-motivated scenario is close to what 
actually happens  
in the decoupling limit  $ m_g \gg \Lambda_{SYM} $. Two different
versions of this scenario may be expected. First, it may happen that
at {\it all} generic values of $ \theta $ there 
exist one true vacuum of lowest energy plus $ (p-1) $ metastable
CP-violating vacua, which are separated by potential barriers and
intertwine when $ \theta $ evolves. Another possibility is that 
metastable vacua exist only in the vicinity of a level crossing 
point in $ \theta $, while for other values of $ \theta $ they 
become the saddle points or maxima \cite{Wit2,Smilga}. 
In one of these forms, 
such a picture seems to be needed to match  
the Witten-Veneziano \cite{WV} resolution of the U(1) problem. 
The scenario discussed in 
\cite{Shifman} implies that the number of vacua $ p $ remains  
$ N_c $ in the limit  $ m_g \gg \Lambda_{SYM}
$, but it is conceivable that an additional level splitting occurs 
with passing the region  $ m_g \sim \Lambda_{SYM}
$ where the SUSY methods become inapplicable. Actually, the 
picture arising in our approach will be just in this vein (see Sect.5).
As will be discussed there, which of the above two versions is 
realized is mostly determined by the value of parameter $ q $. When
$ q \neq 1 $, metastable vacua exist for all
values of $ \theta $, as it happens in the SUSY scenario.
Furthermore, it will be argued that large lifetimes of metastable 
states, necessary for this scenario to work, are ensured 
parametrically - a fact which is not seen 
\cite{Shifman} in the SUSY-motivated 
picture.

One more possible approach we wish to discuss is based on an idea 
formulated some time ago by K\"{u}hn and Zakharov (KZ) \cite{KZ}. 
These authors have suggested that in QCD with massless quarks 
nonperturbative matrix elements should be holomorphic in the 
Pauli-Villars fermion mass $ M_R $. Assuming this kind of holomorphy,
they have proposed to 
relate the proton matrix element of the topological density  
$ \la p | G \tilde{G} | p \ra $ to the matrix element  
$ \la p | G^2 | p \ra $ which is fixed by the conformal anomaly.
However, it is not easy to separate perturbative 
and nonperturbative contributions to the latter due to 
a non-trivial proton wave function. This may be a potentially 
problematic point when the KZ holomorphy is considered for the 
matrix elements\footnote{V.I. Zakharov, private communication.}.
On the other hand, for vacuum condensates and zero momentum 
correlation functions the nonperturbative contributions can 
be systematically singled out, at least formally \cite{2}. 
In this respect, the latter objects are simpler than 
the hadron matrix elements, and thus appear preferable for 
testing the KZ holomorphy. This issue was addressed in \cite{2}
where a method similar to that of Ref.\cite{KZ} was used 
in the context of pure YM theory
to relate the zero momentum two-point function of  
$  G \tilde{G} $ to that of $ G^2 $. In this approach, 
pure gluodynamics was considered as a low energy limit of a 
theory including a heavy quark, while holomorphy in the 
{\it physical} fermion mass  $ 
m \rightarrow \infty $ was argued to hold basing on decoupling 
arguments. Appealing to this holomorphy, it was suggested 
that for the case of pure YM theory the parameters of interest
are $ p = 3 b_{YM}= 11 N_c $ and $ q = 8 $, for 
odd $ N_c $. Furthermore,
as the anomalous conformal WI's for the operator $ G^2 $ 
\cite{NSVZ} {\it are} covariant in $ \theta $, one 
can conclude that the  K\"{u}hn-Zakharov holomorphy, if it holds, 
indeed implies covariance of Eq.(\ref{2a}) in $ \theta $, 
and thus leads to the holomorphic effective potential for 
gluodynamics,
Eq.(\ref{3}). 

An inverse route was undertaken in \cite{3}. In this paper, the 
starting point was the holomorphic effective potential (\ref{3})
for pure YM theory, 
with unspecified parameters $ p $ and $ q $. Again,
the idea was that the meaning of this holomorphy can be clarified 
by coupling pure YM theory to a very heavy fermion with 
mass $ m \rightarrow \infty $, while the values of parameters 
$ p $ and $ q $ would be fixed in this case by some kind of 
consistency conditions. In contrast to the previous approach,
it was suggested in \cite{3} to introduce a heavy fermion 
directly at the effective Lagrangian level by using the 
``integrating in" procedure familiar in the context of SUSY 
theories \cite{Int,SUSY}. Thus, the ``integrating in" method 
is used here to construct an effective Lagrangian for the 
system (YM + heavy fermion) from the effective Lagrangian
for pure YM theory. The consistency condition suggested 
in \cite{3} is that the holomorphic structure of the
latter should arise from the holomorphic structure of the former, 
assuming the standard form of the fermion mass term. Then,
the ``integrating in" method plus the above consistency 
condition are found to select the only possible values 
$ p = 3 b_{YM} = 11 N_c $ and $ q = 8 $. These results 
coincide with the 
ones obtained with the approach of \cite{2}. Thus, the 
``integrating in" method seems to suggest the effective Lagrangian
realization of the K\"{u}hn-Zakharov holomorphy. The agreement 
of these two lines of reasoning is encouraging,
and shows that different assumptions made in \cite{2} and \cite{3}
are at least consistent with each other. 

Finally, we would like to comment on another 
related development.
Very recently, Witten \cite{Wittheta} has shown how the qualitative 
features of the $ \theta $ dependence in non-supersymmetric 
YM theory - such as a multiplicity of vacua $ \sim N_c $,  
existence of domain walls and 
exact vacuum doubling
at some special values of $ \theta $ - can be understood using 
the AdS/CFT duality.  The latter \cite{Mald} 
provides a continuum version of 
the strong coupling limit, with a fixed ultraviolet
cutoff, for YM theory with  
$ N_c \rightarrow \infty $, $g_{YM}^{2} N_c \rightarrow \infty
$. As was shown in \cite{Wittheta},
in this regime the $ \theta $ dependence of the vacuum energy
in YM theory takes the form 
\beq
\label{Wt}
E_{vac} (\theta) = C \, \min_{k} ( \theta + 2 \pi k)^2 + 
O(1/N_c) \; , 
\eeq   
where $ C $ is some constant.
We would like to make two comments on a comparison of
our results  
with the picture advocated by Witten in the large $ N_c $ 
limit. First, we note 
that the structure of Eq.(\ref{Wt}) agrees with our modified 
definition of the path integral including summation over
all branches of a multi-valued (effective) action. Indeed,
Eq.(\ref{Wt}) suggests the correspondence
\beq
\label{Wt2}
C \, \min_{k} \, ( \theta + 2 \pi k)^2  \Leftrightarrow
\lim_{V \rightarrow \infty} \, \left( - \frac{1}{V} \right)
\, \log \left[ \sum_{k} e^{-VC (\theta + 2 \pi k)^2 } \right]
\eeq
using the definition of the vacuum energy through the 
thermodynamic limit of the path integral. With this 
definition which prescribes the way the volume $ V $ appears
in the formula for the vacuum energy, the correspondence
(\ref{Wt2}) appears to be the only possible one. 
On the other hand, the latter expression has exactly  
the structure that arises 
with our definition of the improved effective potential (\ref{3}).
Therefore, our prescription of summation over
all branches of a multi-valued (effective) action
seems to be consistent with the picture developed by Witten
using an approach based on the AdS/CFT correspondence.
In particular, our picture of bubbles of metastable vacua
bounded by domain walls considered in the context 
of QCD in Sect. 5 is in qualitative agreement 
with that suggested by Witten \cite{Wittheta}
for the pure YM case. 

Second, one may wonder whether the approach of Ref. \cite{Wittheta}
can provide an alternative way to fix the parameters $ p,q $ of 
interest. We note that   
Eq.(\ref{Wt}) indicates
a non-analyticity at the 
values $ \theta_c  = \pi \; (mod \, 2 \pi) $ only, where $CP$ 
is broken spontaneously. If the technique based 
on the AdS/CFT duality could be smoothly continued to the weak 
coupling regime of non-supersymmetric YM theory, this would 
result in the values  $ q = 1 , p \sim N_c $. 
However, the possibility of such extrapolation is unclear,
as for small $ \lambda = g_{YM}^{2} N_c $ the 
background geometry develops a singular behavior and the 
supergravity approach breaks down.
There might well be a phase transition \cite{ft}
when the effective YM coupling $ g_{YM}^2 \, N_{c} $ is reduced.
That such a phase transition should occur in 
the supergravity approach to $ QCD_3 $ was argued 
in \cite{GO}. Other reservations about the use of the supergravity
approach to the non-supersymmetric YM theory in D=4 have been expressed
in \cite{Oog} where no perturbative indication was found for 
decoupling of unwanted massive Kaluza-Klein states of string theory.
On the other hand, there exists some evidence from 
lattice simulations that a critical value of $ \theta $
moves from $ \theta_c = \pi $ in the strong coupling 
regime to $ \theta_c < \pi $ in the weak coupling regime \cite{lattice}. 
In terms of parameters $ p ,q $, such a case corresponds to 
$ q \neq 1 $. Therefore, we conclude that if no phase transition
existed in the supergravity approach, the results of Refs. \cite{2,3}
would be 
in conflict with the latter which would imply $ p = O(N_c), q = 1 $.
In this case, the assumptions made in \cite{2,3} would have 
to be reconsidered.
Alternatively, there might be no conflict between the two approaches
if such a phase transition does occur.  
 
To summarize, at the qualitative level we expect a vacuum structure
similar to that suggested by the SUSY scenario. As for the 
quantitative results for the parameters $ p $ and $ q $, different 
lines of reasoning lead to generally different answers. The 
self-duality hypothesis, K\"{u}hn-Zakharov - type arguments and 
non-standard SUSY breaking toy model all suggest $ p = 
3 b_{YM} = 11 N_c $ with 
$ q = 8 $ or 12 for the pure YM case. The 
compatibility of the appealing choice 
$   
G^2 \pm i G \tilde{G} $ , $ p = N_c $ , $ q = 1 $, suggested 
by the soft SUSY breaking scenario, with the 
renormalization group and conformal anomaly for non-supersymmetric 
YM theory remains unclear. It is conceivable that our current 
understanding of the effective Lagrangian is incomplete, and a more 
careful analysis - perhaps 
along the lines of Refs. \cite{Mald,Wittheta} - 
will solve the puzzle (if it {\it is} a puzzle)
of ``extra 1/3 ", thus favoring the SUSY-type scenario. Alternatively,
the numbers  $ p = 11N_c $ , $ q = 8  $ (for odd $ N_c $) suggested 
by the methods of \cite{3,2} may be the correct answer, though 
perhaps somewhat
``counter-intuitive". For these reasons, below we will keep the 
general notation $ p , q $, while a separate analysis will be given
in cases when the concrete values of $ p , q $ are essential. 
The reader is referred to the Appendix for the details of the 
``integrating in" method of \cite{3} adopted to the case of 
full QCD, which suggests the values $ p = 3 b = 11N_c - 2 N_f $ , 
$ q = 8 $.

\section{Effective chiral Lagrangian for finite $ N_c $ }

The anomalous effective potential (\ref{6}) contains both 
the light chiral fields $ U $ and heavy ``glueball" fields
$ h \, , \, \bar{h} $, and is thus not an effective 
potential in Wilsonian sense. On the other hand, only the light  
degrees of freedom, described by the fields $ U $, are relevant 
for the low energy physics. An effective potential for the $ U \, , 
\, \bar{U} $ fields can be obtained by integrating out the 
$ h \, , \, \bar{h} $ fields in Eq.(\ref{6}). It corresponds to 
a potential part of a low energy Wilsonian effective Lagrangian
for energies less than the glueball masses\footnote{
Such an object is not precisely
Wilsonian effective action in the usual
sense as it does not involve e.g. the vector $ \rho, \omega $ mesons 
whose masses are compatible with that of the $ \eta' $.}. The 
transition from 
the effective potential (\ref{6}) for the $ U \, , \,  
h $ fields
to a Wilsonian effective potential for the $ U $ fields
by integrating out the $ h \, , \, \bar{h} $ fields is analogous 
to the transition \cite{Int} from TVY effective Lagrangian 
\cite{TVY} for
SUSY QCD to the Affleck-Dine-Seiberg 
\cite{ADS} low energy effective Lagrangian. The purpose of this section
is to obtain such an effective potential for the light $ U \, , \, 
\bar{U} $ fields by integrating out the $ h \, , \, \bar{h} $ fields
in Eq.(\ref{6}). 

To find this effective potential for the light 
 $ U \, , \, 
\bar{U} $ fields, we make two observations. First, 
the mass term in Eq.(\ref{6}) does not couple to the ``glueball" 
fields, and is thus unessential for integrating them out.
To simplify the subsequent formulas, in this 
section we will omit the mass term, and add it at the end of 
calculation.
Second, the only remaining effect of the light matter fields
 $ U \, , \, 
\bar{U} $ is formally reduced, as was discussed in Sect.2, to 
the redefinition
(shift) (\ref{5}) of the $ \theta $ parameter, and the changes of 
the numerical parameters in the effective potential (\ref{3})
for pure YM theory. Therefore, 
for space -time independent fields  $ U \, , \, 
\bar{U} $ we can integrate out the 
``glueball" fields  $ h \, , \, \bar{h} $ in Eq.(\ref{6})
in the same way as the $ \theta $ dependence of the vacuum
energy for the pure YM case was found in \cite{1}. For the 
sake of completeness, this calculation will be repeated below
for the present case of QCD. As before, we will keep  
 the total space-time 4-volume finite, while a transition to 
the thermodynamic limit $ V \rightarrow
\infty $ will be performed at the very end. 

We start with introducing the ``physical" real fields $ \rho ,\omega $
defined by the relations
\beq
\label{29}
h = 2 E \, e^{ \rho + i \omega} \; \; , \; \; 
\bar{h} = 2 E \, e^{ \rho - i \omega } \; .
\eeq
(This definition implies $ W(\omega + 2 \pi ) = W(\omega) $. As 
will be seen, this condition of single-valuedness of the $ \omega $
field is satisfied with the substitution (\ref{29}).) Then,  
for the ``dynamical" part of Eq.(\ref{6}) we obtain

\beq
\label{30}
- \frac{iV}{4} \left( h \, Log \frac{h}{2 e E} + 
 \bar{h} \, Log \frac{ \bar{h}}{ 2 e E} \right) =
- i V E \, e^{ \rho} \left[ ( \rho -1) \cos \omega - \omega \sin 
\omega 
\right] \; .
\eeq
The summation over the integers $ n $ in Eq.(\ref{6}) enforces 
the quantization rule due to the Poisson formula 
\beq
\label{31} 
\sum_{n} \exp \left(
2 \pi i n  \, \frac{q}{p} \, V \, \frac{h - \bar{h}}{ 4 i} 
\right) = \sum_{n} \delta \left( \frac{q}{p}
\, V E e^{ \rho} \, \sin \omega - n \right) \; ,
\eeq 
which reflects quantization of the topological charge
in the original theory.
Therefore, when the constraint (\ref{31}) is imposed,
Eq.(\ref{30}) can be written as 
\beq
\label{32}
- \frac{iV}{4} \left( h \, Log \frac{h}{2 e E} + 
 \bar{h} \, Log \frac{ \bar{h}}{ 2 e E} \right) =
- i V E \, e^{ \rho} \, ( \rho -1) \cos \omega 
+ i n \, \frac{p}{q} \, \omega \; .
\eeq
Using (\ref{31}),(\ref{32}), we put Eq.(\ref{6}) 
(with the mass term omitted) in the form
\beq
\label{33}
e^{ - i V W} = \sum_{n = - \infty
}^{ + \infty} \sum_{k=0}^{q-1}
 \delta( V E \, \frac{q}{p} \, e^{ \rho} 
\sin \omega - n) \, \exp \left[ - i V E \, e^{ \rho} ( \rho -1)
\cos \omega 
+ i n \left( \bar{\theta}_k +  \frac{p}{q} \, \omega 
\right) \right] 
\eeq
where we denoted
\beq
\label{34}
\bar{\theta}_k \equiv \theta - i \log Det \, U 
+ 2 \pi \, \frac{p}{q} \, k \; .
\eeq
To resolve the constraint imposed by the presence 
of $ \delta $-function in Eqs.(\ref{31}),(\ref{33}),   
we introduce the new field $ \Phi $ by the formula
\beq
\label{35}
 \delta( V E \, \frac{q}{p} \, e^{ \rho} 
\sin \omega - n) \propto \int D \, \Phi \; \exp \left( i \Phi 
V E \, e^{ \rho } \sin \omega - i \Phi 
\, \frac{p}{q} \, n \right) \; 
\eeq
Going over to Euclidean space
 by the substitution 
$ i V \rightarrow V $,
we obtain from Eqs.(\ref{33}),(\ref{35}) 
\bea
\label{36}
W( U , \rho, \omega, \Phi) = - \frac{1}{V} \log \left\{ \sum_{n =
- \infty}^{+ \infty} \sum_{k=0}^{ q-1} \exp 
\left[
- V E e^{ \rho} \left\{ ( \rho - 1) \cos \omega - \Phi
\sin \omega \right\} 
  \right. \right.\nonumber \\ 
+  \left. \left.
i n  \left( \theta 
- i \log Det \, U + 2 \pi k \, \frac{p}{q} + \frac{p}{q} \, \omega - 
\frac{p}{q} \, \Phi \right) -  \varepsilon \, 
\frac{n^2}{ VE} \right] 
\right\} \; .
\eea
Here we introduced the last term to regularize the infinite sum over  
the integers $ n $. The limit $ \varepsilon \rightarrow 0 $ will be 
carried out at the end, but before taking the thermodynamic limit 
$ V \rightarrow \infty $. Note that Eq.(\ref{36}) satisfies the 
condition $ W (\omega + 2 \pi) = W (\omega) $ which should hold as 
long as $ \omega $ is an angle variable. We also note that the 
periodicity in $ \theta $ with period $ 2 \pi $ is explicit in
Eq.(\ref{36}).
 
To discuss the thermodynamic limit $ V \rightarrow \infty $ we 
use the identity 
\beq
\label{a7}
 \theta_3(\nu, x)=\frac{1}{\sqrt{\pi x}}\sum_{k=-\infty}^{+\infty}
\exp \left[ -\frac{(\nu+k)^2}{x} \right] =
  \sum_{l=-\infty}^{+\infty}
 \exp [ -l^2\pi^2x +2il\nu\pi ]
\eeq
and transform Eq.(\ref{36})
into its dual form
\bea
\label{38}
W (U,  \rho, \omega, \Phi) = - \frac{1}{V} \log \left\{ \sum_{n =
- \infty}^{ + \infty} \sum_{k=0}^{ q-1} \exp 
\left[
- V E e^{ \rho} \left\{ ( \rho - 1) \cos \omega -
 \Phi \sin \omega \right\} 
  \right. \right. \nonumber \\ 
-  \left. \left.
\frac{VE}{ 4 \varepsilon} 
\left( \theta - i \log Det \, U +  2 \pi k \, \frac{p}{q} + 
\frac{p}{q} \, 
\omega - 
\frac{p}{q} \, \Phi - 2 \pi n \right)^2  \; \right] \right\} \; ,
\eea
where we have omitted an irrelevant infinite factor  
$ \sim \varepsilon^{-1/2} $
in front
of the sum. 
Eq.(\ref{38}) is the final form of the improved effective potential
$ W $, which is suitable for integrating out the 
``glueball" fields $ \rho \, , \, \omega $, along with the 
auxiliary 
field $ \Phi $. 
 To this end, the function $ W $ 
should be minimized in respect to the three variables 
$ \rho , \omega $ and $ \Phi $, with $ U $ fixed. In spite 
of the frightening form
of this function, its extrema can be readily found using the 
following 
simple trick. As at the extremal points all partial derivatives 
of the 
function $ W $ vanish, we first consider their linear combination
in which the sum over $ n , k $ cancels out. We thus arrive at the 
equations
\bea
\label{39}
\frac{ \partial W}{ \partial \rho} =  E e^{ \rho}
( \rho \cos \omega - \Phi \sin \omega ) &=& 0 \nonumber \\
\frac{ \partial W}{ \partial \omega} + 
\frac{ \partial W}{ \partial \Phi} = - E e^{ \rho} (
\rho \sin \omega + \Phi \cos \omega ) &=& 0 \; ,
\eea
which is equivalent to $ \rho^2 + \Phi^2 = 0 $. (We do not consider
here the case $ \rho \rightarrow - \infty $ which would 
also solve Eqs. (\ref{39}), see \cite{3}.) Therefore,
these equations have the only solution 
\beq
\label{40}
\la \rho \ra = 0 \; \; , \; \; \la \Phi \ra = 0 \; , 
\eeq
while the minimum value of the 
angular field $ \omega $ is left arbitrary by them.
The latter can now be found from either of the 
constraints $ 
\partial W / \partial \omega = 0 $ or 
$ \partial W / \partial \Phi = 0 $, which become identical for $ 
\la \rho \ra  = \la \Phi \ra = 0 $. The resulting equation 
reads
\bea
\label{41} 
\sum_{n =
- \infty}^{+ \infty} \sum_{k=0}^{q-1} 
\left( \theta  - i \log Det \, U + 2 \pi k \, \frac{p}{q} - 
2 \pi n + \frac{p}{q} \, 
\omega +  2 \varepsilon \, \frac{q}{p} \sin \omega \right) 
\nonumber \\
\times \exp \left\{ V E \cos \omega  
 - \frac{VE }{ 4 \varepsilon}
\left( \theta  - i \log Det \, U + 
2 \pi k \, \frac{p}{q}  + \frac{p}{q} \, 
\omega  - 2 \pi n \right)^2 \right\} = 0 \; ,
\eea
in which we have to take the limit $ \varepsilon \rightarrow 0 $ 
at a fixed 4-volume $ V $. 
One can see that non-trivial solutions of Eq.(\ref{41}) at 
$ \varepsilon \rightarrow 0 $ are given by 
\beq
\label{42}
\la \omega \ra_l = - \frac{q}{p} \, 
( \theta  - i \log Det \, U ) + \frac{2 \pi}{ p}
\, l + 2 \pi r \; \; , \; \; l = 0,1, \ldots , p-1 \; \; ; 
\; \; r = 0, \pm 1, \ldots
\eeq
Eq.(\ref{42}) shows that  $ p $ 
physically distinct solutions of the equation of motion for the 
$ \omega $ field, while 
the series over the integers $ r $ in Eq.(\ref{42}) simply 
reflects the
angular character of the $ \omega $ variable, 
and is thus irrelevant.
Substituting  Eq.(\ref{42}) back to Eq.(\ref{38})
and restoring the mass term for the $ U $ field, we 
obtain the effective potential for 
the light chiral fields  $ U \, , \, 
\bar{U} $ \cite{QCD}:
\bea
\label{8}
W_{eff}(U,U^{+}) =  - \lim_{V \rightarrow 
\infty} \; \frac{1}{V} \log \left\{ 
 \sum_{l=0}^{p-1} \exp \left[ 
V E \cos \left[ - \frac{q}{p} ( \theta - i \log  Det \, U ) 
+ \frac{2 \pi}{p}
\, l \right]  \right. \right. \nonumber \\
\left. \left. + 
\frac{1}{2} V \, Tr ( M U + M^{+} U^{+} ) \right] \right\}   
\; . 
\eea
Eq.(\ref{8}) is the final form of the effective potential for 
the chiral field which is valid for any value of $ \theta $. 
As was mentioned in the beginning of this section, this
form of the effective potential (\ref{8}) could be read off 
the formula \cite{1} for the vacuum energy as a function of 
$ \theta $ in pure YM theory, with the substitution of 
the parameters $ p,q $ by their values in QCD, adding the mass 
term for the chiral field, and making the shift (\ref{5}) 
of the $ \theta $ parameter. The meaning of the 
resulting expression is different, however. What was the 
the vacuum energy as a function of $ \theta $ describing 
$ p $ different vacua in YM theory 
becomes the effective potential for the light chiral fields. 
At first sight, it could be expected that the resulting effective
chiral Lagrangian has the same number $ p $ of different vacua. 
It turns out that this naive expectation is wrong: the number 
of different vacua in QCD is determined by the number of 
flavors $ N_f $, at least as long as $ N_f < N_c $. The reason why 
the number of vacua differ from the number of branches of the 
effective potential is the angular character of the corresponding 
chiral degrees of freedom. We hope that the last sentence will 
become clearer in the next section when we consider the concrete
examples. 

The peculiarity of the resulting effective potential (\ref{8}) is 
that
it is impossible to represent it by a single analytic
function by directly
performing the limit $ V \rightarrow \infty $ in (\ref{8}).
In the thermodynamic limit   $ V \rightarrow \infty $ the 
only surviving term in the sum in Eq.(\ref{8}) is the one
maximizing the cosine function. Thus, the thermodynamic limit 
selects, for given value of $ \theta -  i \log  Det \, U  $, a 
corresponding value 
of $ l $, i.e. one particular branch in Eq.(\ref{8}). 
The branch structure of Eq.(\ref{8}) shows up in the limit 
$ V \rightarrow \infty $ by the presence of cusp singularities 
at certain values of  $ \theta -  i \log  Det \, U  $.
These cusp singularities are analogous to the ones arising in 
the case of pure gluodynamics \cite{1} for the vacuum energy as 
function of $ \theta $, showing non-analyticity
of the $ \theta $ dependence at certain values of $ \theta $.
In the present case, the effective potential for the light 
chiral fields analogously becomes non-analytic at some 
values of the fields. The origin of this non-analyticity 
is the same as in the pure YM case - it appears when the topological
charge quantization is imposed explicitly at the effective 
Lagrangian level.
 
The general analysis
of the effective potential (\ref{8}) will be given in the 
next section, while here we consider the case when the combination
$ \theta -
i \log Det \, U $ is small, and thus   
the term with $ l = 0 $ dominates. 
We obtain for this case
\beq
\label{9}
W_{eff}^{(l=0)}( U, U^{+}) = - E \cos \left[ - \frac{q}{p} ( \theta -
i \log Det \, U ) \right] - \frac{1}{2} \,  
Tr \, (MU + M^{+}U^{+} ) \; .
\eeq
Expanding the cosine (this corresponds to the expansion
in $ q/p \sim 1/N_c $), we recover
exactly  the ECL of \cite{Wit2} at lowest order 
in $ 1/N_c $ (but only for small $ \theta - i \log Det \, U
< \pi/ q $), together with 
the ``cosmological" term $ - E = - 
 \la b \alpha_s /(32 \pi) G^2 \ra $ required by the conformal 
anomaly:
\beq
\label{VVW}
W_{eff}^{(l=0)}( U, U^{+}) = - E - \frac{ \la \nu^2 \ra_{YM} }{2} 
( \theta - i \log Det \, U )^2 - 
\frac{1}{2} \, Tr \, (MU + M^{+}U^{+} ) + \ldots \; , 
\eeq
where we used the fact that, according
to Eq.(\ref{2a}), at large $ N_c $ $  E(q/p)^2 = - 
\la \nu^2 \ra_{YM} $ where $ \la \nu^2 \ra_{YM}< 0 $ is the 
topological susceptibility in pure YM theory. 
Corrections in $ 1/N_c $ stemming from Eq.(\ref{9}) 
constitute a new result.
Thus, in the large $ N_c $ limit 
the effective chiral potential (\ref{8}) coincides 
with that of \cite{Wit2} in the vicinity of the global minimum.
At the same time, terms with $ l \neq 0 $ in Eq.(\ref{8}) result 
in different global properties of the effective chiral potential
in both cases $ q = 1 $ and $ q \neq 1 $
in comparison with the one of Ref.\cite{Wit2}, see below.

\section{Theta dependence, metastable vacua and domain walls}

In this section we analyse the 
picture of the 
physical $ \theta $ dependence and vacuum structure
stemming  
from the effective potential (\ref{8}). This is where we 
encounter the main difference 
of our results from the scenario of \cite{Wit2}. The origin 
of this difference is the branch structure of the 
effective potential (\ref{8}), with 
the prescription of summation over all branches.
As we have mentioned earlier, this effective potential 
has cusp singularities at certain values of the fields,
whose origin is the topological charge quantization
in the effective Lagrangian framework.
It is therefore clear that these cusp singularities 
can not be seen in the usual treatment of the 
effective chiral Lagrangian, which deals 
from the very beginning with quark degrees 
of freedom only without imposing quantization of the 
topological charge. 
Thus, the different global form of the 
effective chiral potential as a function of the chiral condensate
phases $ \phi_i $, with cusp singularities at certain values of 
the phases $ \phi_i $, is the first essential difference of 
our picture
from that of Ref.\cite{Wit2}. Another important difference 
appears when
$ q \neq 1 $. In this case, we will find metastable vacua, 
separated 
by the barriers from the true physical vacuum of lowest energy, 
in the whole range of variation of $ \theta $. Properties
of these metastable vacua will be discussed below.
Furthermore, 
for the same case $ q \neq 1 $ we will argue that 
the vacuum doubling at the
points $ \theta = (2 k +1) \pi/ q $ 
occurs irrespective of the particular values of the 
light quark masses (no Dashen's constraint, see below).

It is convenient to describe the non-analytic effective
chiral potential (\ref{8}) by a set of analytic functions
defined on different intervals of the combination
$ \theta - i \log Det \; U $. Thus,  
according to Eq.(\ref{8}), 
the infinite volume limit of the effective potential for the 
fields $ U= diag( \exp i \phi_q ) $ 
is dominated by its $l $-th branch: 
\beq
\label{13}
W_{eff}^{(l)} = - E \cos \left( - \frac{q}{p} \theta + 
\frac{q}{p} \sum \phi_{i} + \frac{2 \pi}{p} \, l \right) 
- \sum M_{i} \cos \phi_i  \; \; , \; \;   l= 0,1, \ldots , p-1
\eeq
if 
\beq
\label{14}
(2 l - 1) \frac{\pi}{q} \leq \theta - \sum \phi_i < (2 l + 1) 
 \frac{\pi}{q} \; \;  .
\eeq
This can be viewed as the set of $ ``p" $ different effective 
potentials 
describing different branches in Eq.(\ref{6}). The 
periodicity in $ \theta $ is realized on the set of 
potentials (\ref{13}) as a whole, precisely as it occurs in the 
pure gauge case \cite{1} where different branches undergo a
cyclic permutation under the shift $ \theta \rightarrow 
\theta - 2 \pi $. 
As is seen from Eq.(\ref{13}), the shift 
 $ \theta \rightarrow \theta - 2 \pi $ transforms the branch with
$ l = k $ into the branch with $ l = k + q $. 
In addition, as long as $ q \neq 1 $, there exists another 
series of cyclic permutations
corresponding to $ l = k $ and $ l = k+1 $ in the above 
set, which are related to each 
other by the shift $ \theta \rightarrow \theta - 2 \pi/q $.
If the number $ q $ were $ q = 1 $, the 
two series would be, of course, the same. As was mentioned in Sect.2,
a value $ q \neq 1 $ thus implies some discrete symmetry arising
at the quantum level.
(Although the periodicity in $ \theta $ with period $ 2 \pi/q $, 
$ q \neq 1$,   
rather than $ 2 \pi $ may look surprising, it is not 
the first time we encounter such a situation. We would 
like to note in $ N = 2 $ Seiberg-Witten theories with 
quarks the $ \theta $ dependence for $ N_c = 2 $ has period 
$ \pi $, and not the ``standard" $ 2 \pi $ \cite{Konishi}.
A similar behavior was argued to hold in some 2D models on the 
lattice \cite{lattice}.) 
In what follows we will discuss both cases 
$ q = 1 $ and $ q \neq 1 $.  

\begin{figure}
\epsfysize=3in
\epsfbox[-40 275 508 686]{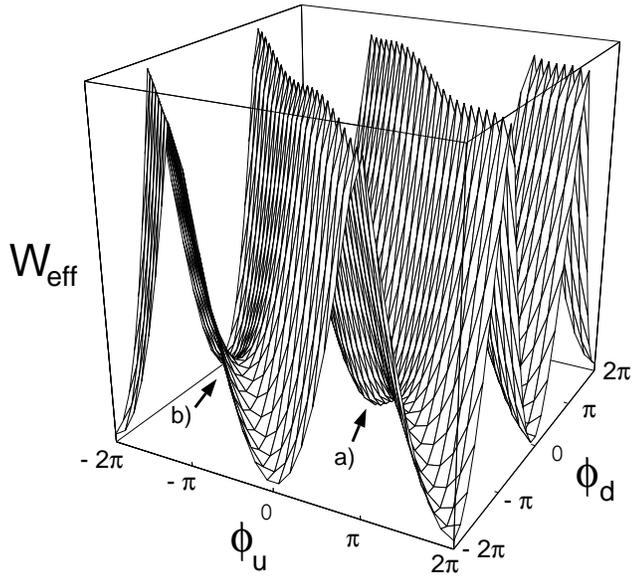}
\caption{Two flavor effective potential for $ q = 1 $.  
Indicated points are 
a) minimum and
b) saddle point.}
\label{twoflavor}
\end{figure}

Consider the equation of motion for the lowest branch $ l = 0 $ :
\beq
\label{15} 
\sin \left( \frac{q}{p} \theta - \frac{q}{p} \sum \phi_i 
\right) = \frac{p}{q} \frac{M_i}{E} \, \sin \phi_i \; \; , \; \; 
i = 1,\ldots, N_f
\eeq 
with the constraint (\ref{14}) with $ l = 0 $. At lowest order in 
$ 1/N_c $ this equation coincides with that of \cite{Wit2}. For 
general values of $ M_{i} / E $, it is not possible to solve 
Eq.(\ref{15}) analytically.
However, in the realistic case $ \varepsilon_{u},
\varepsilon_{d} \ll 1 \, , \, \varepsilon_s \sim 1 $ where 
$ \varepsilon_{i} = (p/q) M_{i}/ E $, the approximate
solution can be found. Neglecting the
$ O(\varepsilon_{u}, \varepsilon_{d}) $ terms in the phases 
$ \phi_i $ 
(in this approximation we deal with the case $ N_f = 2 $),
we obtain
\bea
\label{20}
\phi_{s}^{(l=0)} &=& 0 \nonumber \\
\phi_{u}^{(l=0)} + \phi_{d}^{(l=0)} &=& \theta  \\
\varepsilon_{u} \sin \phi_{u}^{(l=0)} &=& 
\varepsilon_{d} \sin \phi_{d}^{(l=0)} \nonumber 
\eea
One sees that the constraint (\ref{14}) is automatically satisfied.
The solution of Eqs.(\ref{20}) reads
\bea
\label{21}
\sin \phi_{u}^{(l=0)} &=& 
\frac{ m_d \sin \theta }{ [m_{u}^2 + m_{d}^2 +
2 m_{u} m_{d} \cos \theta ]^{1/2} } + O(\varepsilon_{u},
\varepsilon_{d}) \; , \nonumber \\
\sin \phi_{d}^{(l=0)} &=& 
\frac{ m_u \sin \theta }{ [m_{u}^2 + m_{d}^2 +
2 m_{u} m_{d} \cos \theta ]^{1/2} } + O(\varepsilon_{u},
\varepsilon_{d}) \; ,  \\
\sin \phi_{s}^{(l=0)} &=&  O(\varepsilon_{u},
\varepsilon_{d})  \nonumber \; .
\eea
Thus, the solution for the $ l = 0 $ branch coincides with the 
one of Ref.\cite{Wit2} to the leading order in $ 
\varepsilon_{u},
\varepsilon_{d} $. Let us now concentrate on the case $ q = 1 $. 
For the next $ l = 1 $ branch we obtain, instead of (\ref{20}),
\bea
\label{20a}
\phi_{s}^{(l=1)} &=& 0 \nonumber \\
\phi_{u}^{(l=1)} + \phi_{d}^{(l=1)} &=& \theta - 2 \pi \\
\varepsilon_{u} \sin \phi_{u}^{(l=1)} &=& 
\varepsilon_{d} \sin \phi_{d}^{(l=1)} \nonumber 
\eea
One can easily see that the solution $ \phi_{i}^{(l=1)} $
of these equations can be obtained from the previous one:
$ \phi_{u,d}^{(l=1)} =  \phi_{u,d}^{(l=0)} - \pi $. 
Obviously, solutions for branches with $ l > 1 $ will coincide with  
one of the 
two solutions   $ \phi_{i}^{(l=0)},   \phi_{i}^{(l=1)} $
modulo $ 2 \pi $. Furthermore, while the 
first solution  $ \phi_{u,d}^{(l=0)}  $
defines the location of the global minimum of the effective 
potential, the additional
solution  $ \phi_{u,d}^{(l=1)} $
is the saddle point, see Fig. \ref{twoflavor} for the form of 
the effective potential
at $ \theta = 0 $ for $ N_f =2 $ and $ q = 1 $. 
Such a saddle point of the effective chiral potential may be of 
importance for cosmology and/or the physics of heavy ion collisions.
We will not discuss these issues in this paper, but hope to return to
them elsewhere.

We have checked numerically that in the case $ q = 1 $,
$ N_f = 3 $ with physical
values of the quark masses we stay with one physical vacuum at all 
values of $ \theta $. In particular, no (stable or metastable) vacua
appear at $ \theta = \pi $. Thus, for $ q = 1 $ 
the counting of vacua in our approach
agrees with that of Ref.\cite{Wit2}. On the other hand, when the 
Dashen's constraint \cite{Dash}
\beq
\label{16}
m_{u} m_{d} > m_{s} | m_{d} - m_{u} | \; 
\eeq
is satisfied, metastable vacua appear in the vicinity of the 
point $ \theta = \pi $. In particular, in the $ SU(2) $ limit 
$ m_u = m_d \neq m_s $ the metastable vacuum exists in a very narrow
region near $ \theta = \pi $. When $ \theta $ becomes exactly 
equal $ \pi 
$, the two vacua are exactly degenerate. This agrees with 
the picture of Ref.\cite{Wit2}. 
 
\begin{figure}
\epsfysize=3in
\epsfbox[30 432 498 716]{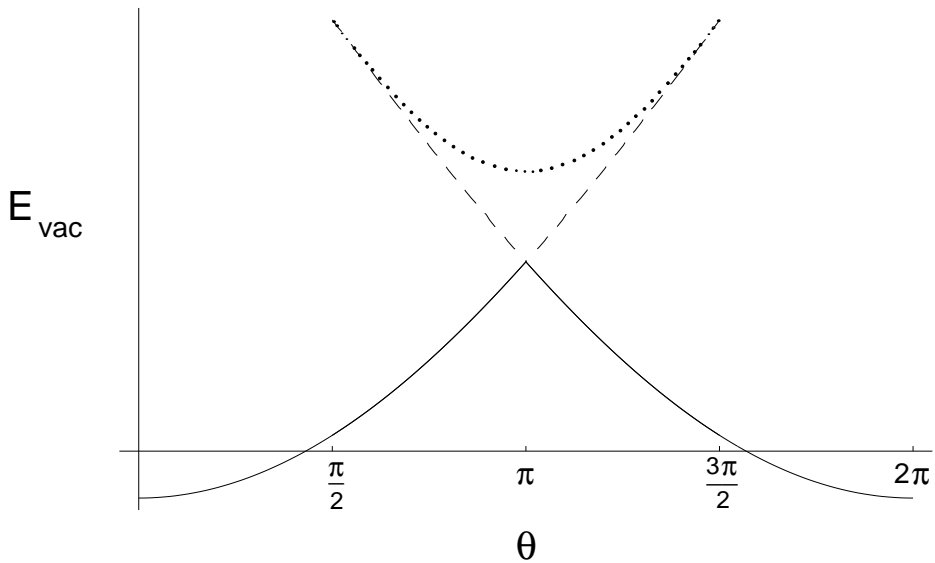}
\caption{Vacuum energy as a function of vacuum angle
$ \theta $ for the case $ q = 1 $ with $ N_f = 3 $
light flavors of equal masses.  The solid and
dashed lines represent minima and the dotted line is the 
intermediate saddle
point.}
\label{thepotential}
\end{figure}

Finally, we consider the case $ q = 1 $ with $ N_f = 3 $ light 
flavors of equal masses. In this case, the metastable vacua 
exist in the extended region of $ \theta $ from $ \pi/2 $ to 
$ 3 \pi/2 $, analogously to what was recently found by 
Smilga \cite{Smilga} for the VVW potential in the same limit.
The resulting picture of  the $ \theta $ dependence of the 
vacuum energy
is shown in Fig. \ref{thepotential}. 

Let us now consider the case $ q \neq 1 $. For the simplest 
situation of isospin $ SU(N_f) $ symmetry   
with masses $ m_{i} = m \ll \Lambda_{QCD} $, the 
lowest energy state is described by
\beq
\label{18}
\phi_{i}^{(l=0)} = \frac{\theta}{N_{f}} \; , \;   E_{vac} 
(\theta) = - E - M  N_{f}
\cos \left( \frac{\theta}{N_{f}} \right) + O(m_{q}^2)  \; , 
\; \theta
\leq \frac{\pi}{q} \; ;  
\eeq
\beq
\label{19}
\phi_{i}^{(l=1)} = \frac{\theta}{N_{f}} - \frac{2 \pi}{q N_f}
\; , \;   E_{vac} (\theta) =  - E - M N_{f} 
\cos \left( \frac{\theta}{N_{f}}- \frac{2 \pi}{q N_f} 
\right) + O(m_{q}^2)  \; , \; \frac{\pi}{q} \leq \theta
\leq \frac{3 \pi}{q} \; ; 
\eeq
{\it etc}. Thus, the solution (\ref{18}) coincides with the 
one obtained 
by VVW \cite{Wit2} at small $ \theta < \pi/ q $ up to $ O(m_{q}^2) $ 
terms. 
However, at larger values of $ \theta $ the true vacuum switches 
from (\ref{18}) to (\ref{19}) with a cusp singularity developing at 
$ \theta = \pi/q $. Here we remind the reader that the ``standard"
location of the first critical point $ \theta = \pi $ corresponds
to the particular case $ q = 1 $ in our general formulas.  
On the contrary, in the scenario \cite{Wit2}
the solution remains, if $ q \neq 1 $, analytic at this point, and 
for $ \pi/q \leq \theta \leq 3 \pi/ q $ has an energy larger 
than in (\ref{19}).
On the other hand, if $ q = 1 $, the picture of Ref.\cite{Wit2}
is reproduced. Moreover, the number of different 
solutions (which may or may not be metastable vacua, depending 
on the signs of second derivatives of the potential) 
is precisely 
$ N_f $ (or $ q N_f $ if $ q \neq 1 $) 
as the phases are defined modulo $ 2 \pi $, and 
thus only the first $ N_f $ ($q N_f $) terms 
in the series (\ref{18},\ref{19})
become operative.
 
\begin{figure}
\epsfysize=3.6in
\epsfbox[-80 263 551 716]{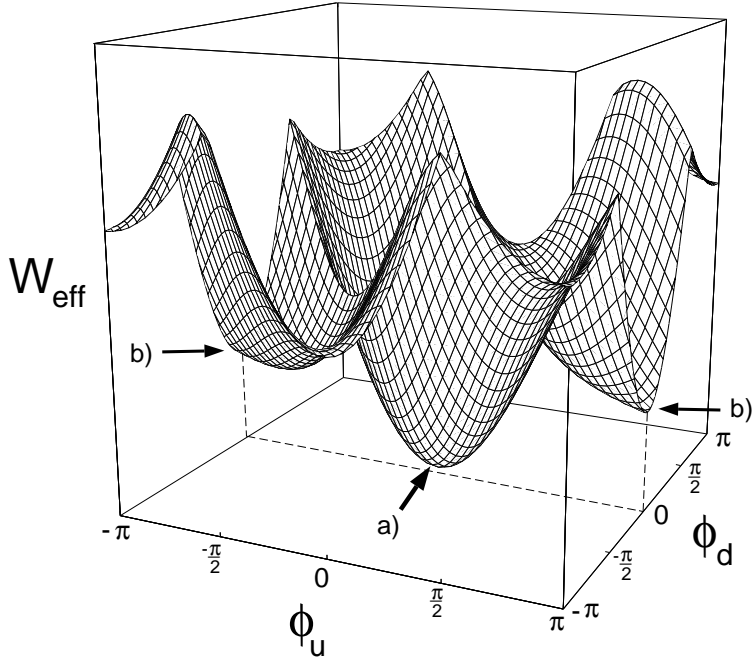}
\caption{Effective potential for q=2 and $ N_f = 2 $. 
The point indicated as a) is the global minimum. 
The two points indicated as b) are identified.}
\label{q=2}
\end{figure}

 The interesting feature of the case $ q \neq 1 $ is 
that the vacuum doubling at the points
\beq
\label{17}
\theta_k = (2 k + 1) \, \frac{\pi}{q} \; \; , \; \;
 k = 0, 1, \ldots, 
p-1 
\eeq
holds irrespective of the values of the light quark masses.
This can be seen from the 
fact that
the equations of motion for any two branches with 
$ l = k $ and $ l = k+1 $ from the set (\ref{13}) are related 
by the shift $ \theta \rightarrow \theta - 2 \pi/ q $. 
Thus, the 
extreme sensitivity of the theory to the values of the light
quark masses in the vicinity
of the critical point in $ \theta $ is avoided in our scenario
if $ q \neq 1 $, while the location of the critical point is 
given by $ \theta_c = \pi/q $ instead of 
the ``standard" $ \theta_c =
\pi $. (A similar situation was argued to hold in 2D 
$ CP^N $ models
on the lattice in the weak coupling limit \cite{lattice}.)
Another interesting feature of the scenario $ q \neq 1 $ 
is the appearance of metastable vacua which exist for any value 
of $ \theta $, including $ \theta = 0 $. For the physical 
values of the quark masses, we find $ q - 1 $ additional local
minima of the effective chiral potential, which are separated 
by barriers from the true physical vacuum of lowest energy. 
For the illustrative purpose, we present in Fig. \ref{q=2} the 
effective potential at $ \theta = 0 $ for $ q = 2 , N_f = 2 $.

The existence of additional local minima of an effective potential
for the case $ q \neq 1 $ leads to the well known phenomenon of 
the false vacuum decay \cite{Okun}. For the effective 
chiral potential
(\ref{8}) this effect and its possible consequences in the 
context of  
axion physics were briefly discussed in \cite{axion}. Below 
we present a somewhat more detailed discussion of this issue.

\begin{figure}
\epsfysize=3in
\epsfbox[ -40 416 604 772]{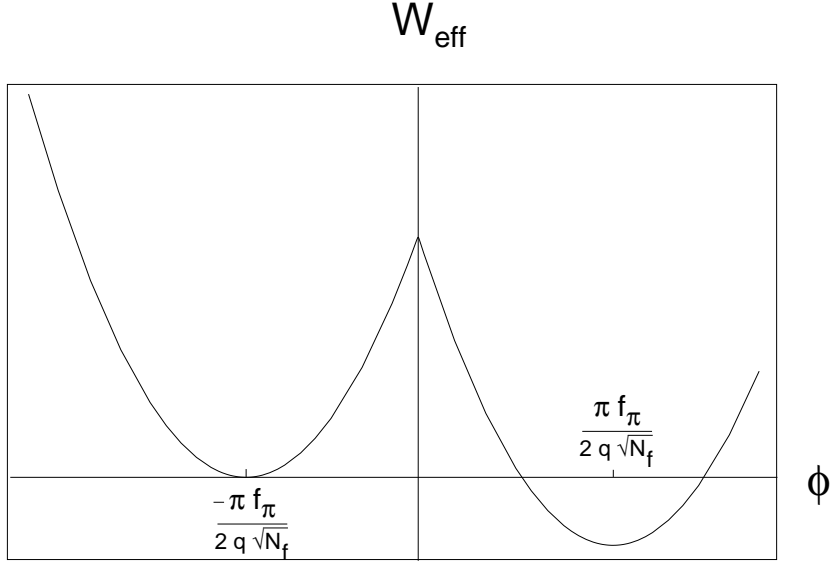}
\caption{Effective potential
at $ \theta = 0 $ for equal chiral phases 
  $\phi=\phi_u=\phi_d=\phi_s$.}
\label{tunneling}
\end{figure}

We discuss the problem in a simplified setting by considering the 
isospin $ SU(N_f) $ limit with equal (and small) fermion masses and 
taking all chiral phases $ \phi_i $ equal, $ \phi_i = \phi $, 
i.e. restricting our analysis
to the ``radial" motion in the $ \phi $-space. In this setting, the 
problem becomes tractable in the spirit of Ref.\cite{Okun}. In 
what follows, we only consider  
transitions between a metastable state of lowest energy and the vacuum.
To calculate the wall surface tension $ \sigma $, it is convenient 
to shift the vacuum energy by an overall constant such that the 
metastable state has zero energy, and to rescale and shift the 
chiral field $ \phi \rightarrow (2 / f_{\pi} \sqrt{N_f} ) \phi - 
\pi/(q N_f ) $ in 
order to have the standard normalization of the kinetic term
and symmetrized form of the potential. 
With these conventions, the effective potential for $ \theta = 0 $
becomes 
\bea
\label{18a}
W (\phi) = \left\{ \begin{array}{ll}
E \left[ 1 - \cos \left( \frac{ 2 q \sqrt{N_f}}{p f_{\pi}} \phi 
- \frac{ \pi}{p} \right) \right]
-  M f(\phi)  
  & \mbox{if $ \phi \geq 0   $ } \\ \nonumber 
E \left[ 1 - \cos \left( \frac{ 2 q \sqrt{N_f}}{p f_{\pi}} \phi
+ \frac{ \pi}{p} \right)  \right]
-  M f(\phi)
& \mbox{if $ \phi \leq  0   $ } 
\end{array} 
\right. \nonumber \\
f(\phi) =  N_f \left[ \cos \left( \frac{2}{f_{\pi} \sqrt{N_f}} \phi - 
 \frac{ \pi}{q N_f } \right) - \cos \left( \frac{2 \pi}{q N_f} 
\right) \right]
\eea
The effective potential (\ref{18a}) has a global minimum
at $ \phi =  \pi f_{\pi}/(2 q 
\sqrt{N_f} ) $ and a local minimum at $ \phi = - \pi f_{\pi}/(2 q 
\sqrt{N_f} ) $, with a cusp singularity between them (see 
Fig. \ref{tunneling}). We note that analogous ``glued" potentials 
were discussed for SUSY models in a similar context \cite{KS,KSS,KKS}.

A few comments on the above effective potential are in 
order. First, we note that the potential barrier is high
($ \sim \la G^2 \ra$ ) and wide, while the energy splitting
for $ \theta = 0 $
 \beq
\label{200}
 \Delta E =   m_q N_f \left| \la \bar{\Psi}\Psi\ra \right| 
\left( 1-\cos  \frac{2\pi}{qN_{f}} \right)  +0(m_q^2),
\eeq 
is numerically small in comparison to the leading term
proportional to the gluon condensate. Therefore, the thin wall
approximation of Ref.\cite{Okun} is justified for the physical
values of the parameters that enter the effective potential.
Indeed, it is easy to check that the necessary condition 
\cite{Okun} $ 3 \sigma \mu \gg  \Delta E $ (where $ \sigma $ 
is the wall surface tension given by Eq.(\ref{19d})
below, and $ \mu $ is the width of the wall)
is satisfied to a good accuracy for both alternative choices 
$ p = 11 N_c - 2 N_f = 27 \, , \, q = 8 $ or $ p = N_c = 3 \, , 
\, q = 1 $. (In the latter case, a metastable state does not exist
for $ N_f = 3 $ and $ \theta = 0 $, 
but is possible for, say, $ N_f = 5 $.) Second, we would like to 
comment on the meaning of the cusp singularity of the effective 
potential (\ref{18a}). As was mentioned earlier, the cusp arises
as a result of integrating out the ``glueball" degrees of freedom,
which were carrying information on the topological charge 
quantization. For an analogous situation in the supersymmetric 
case, it was argued \cite{KKS} that the cusp, where the adiabatic 
approximation breaks down, provides a leading contribution
to the wall surface tension. In our case, we expect the 
difference of the surface tensions for the potential (\ref{18a})
and a potential where the cusp is smoothed to be down 
by powers of $ N_c $. The reason is that the domain wall to be 
discussed below is in fact the $ \eta' $ wall, while on the 
other hand, a coupling of the $ \eta' $ to the glueball fields
near the cusp would yield the above suppression. 

\begin{figure}
\epsfysize=3in
\epsfbox[-50 414 572 782]{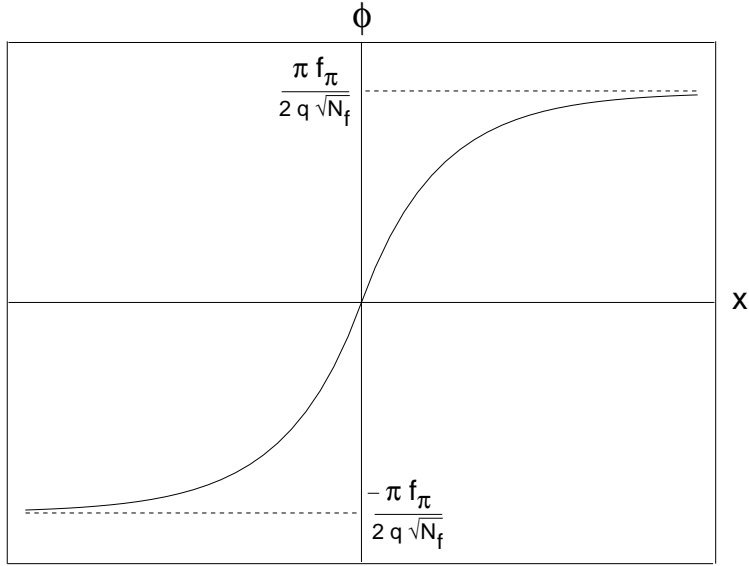}
\caption{Domain wall profile.} 
\label{tunneling_solution}
\end{figure}

Explicitly, the domain wall solution 
corresponding to the effective potential 
(\ref{18a}) is 
\bea
\label{19a}
\phi (x) &=& \frac{p f_{\pi} }{2 q \sqrt{N_f}} \left[ - \frac{\pi}{
p} + 4 \arctan \left\{ \tan \left( \frac{\pi}{4 p} \right) \exp [
\mu (x - x_0)] \right\}  \right] \; \; , \; \; x < x_0  
\nonumber \\
\phi (x) &=& \frac{p f_{\pi}}{2 q \sqrt{N_f}} \left[  \frac{\pi}{
p} - 4 \arctan \left\{ \tan \left( \frac{\pi}{4 p} \right) 
\exp \left[
- \mu (x - x_0) \right] \right\}  \right] \; \; , \; \; x > x_0 \; ,  
\eea 
where $ x_0 $ is the position of the center of the domain wall and 
\beq
\label{19c}
\mu \equiv \frac{2 q \sqrt{N_f} \sqrt{E}}{ p f_{\pi} }  
\eeq 
is the width of the wall, which turns out to be exactly equal to
the $ \eta' $ mass in the chiral limit,  
see Eq. (\ref{12}) below. This suggests the interpretation of the 
domain wall (\ref{19a})  
as the $ \eta' $ domain wall.
The solution (\ref{19a}) as a function of $ x - x_0 $ 
is shown in Fig. (\ref{tunneling_solution}).
Its first derivative is continuous at $ x = x_0 $, but the 
second derivative exhibits a jump.

The wall surface tension
can be easily calculated from Eq.(\ref{18a}). For $\theta=0$ it
is 
\beq
\label{19d}
\sigma=\frac{4 p}{q \sqrt{N_f} }  
f_{\pi}\sqrt{\la \frac{ b \alpha_s }{ 32 \pi} G^2 \ra }
\, \left( 1 - \cos \frac{\pi}{ 2 p} \right) +0(m_q f_{\pi}^2 ).
\eeq 
Eq.(\ref{19d}) should be compared with the formula
 \beq
\label{19e}
\sigma= 3\sqrt{2} \left(1-\frac{\pi}{3\sqrt{3}}
\right) m_{\pi}f_{\pi}^2 .
\eeq 
found by Smilga \cite{Smilga} for the 
the wall surface tension 
at $ \theta \simeq \pi $ for the VVW potential \cite{Wit2} with 
$ N_f = 3 $ with equal fermion masses.
A distinct difference between  these two cases is the absence
of the chiral suppression $\sim m_q$ in Eq.(\ref{19d}),
which apparently 
would make penetration through the barrier even more difficult
in comparison to the VVW potential. Another difference is the large 
$ N_c $ behavior of Eqs. (\ref{19d}) and (\ref{19e}). For fixed
$ m_q \neq 0 $, the surface tension scales as $ O(N_c ) $ for both 
cases, while in the limit $ m_q \rightarrow 0 $ first and then
 $ N_c \rightarrow \infty $ the surface tension vanishes for Eq.
(\ref{19e}) and scales as $ N_{c}^{1/2} $ for Eq. (\ref{19d}). 
  
The quasiclassical formula 
for the decay rate per 
unit time per unit volume is \cite{Okun}
 \beq
\label{19f}
\Gamma\propto\exp \left( -\frac{27\pi^2\sigma^4}{2(\Delta E)^3} 
\right) 
\equiv \exp ( - S_4 ) .
\eeq 
Using (\ref{19e}), one finds that in the VVW scenario the lifetime
of the metastable state at zero temperature
is much larger than the
age of the Universe \cite{Smilga}. In our case, we find a very
different result\footnote{The $ N_c $ dependence displayed in 
Eq.(\ref{19g}) may look suspect as it apparently indicates
that $ \Gamma = O( e^{-1/N_c}) $ as $ N_{c} \rightarrow \infty $.
Such a conclusion would be wrong,
as in Eq.(\ref{19d}) we have neglected the second term in the  
effective potential (\ref{18a})
in comparison to the first one. However, from the point of view 
of the $ N_c $ counting, the second term in Eq.(\ref{18a}) is 
just the leading one. Therefore, it would be erroneous to
extrapolate Eq.(\ref{19g}) to the limit of very large  
$ N_c $. One can easily check that in the limit $ N_c \rightarrow 
\infty $ with $ m_q \neq 0 $ fixed, the lifetime of a metastable 
vacuum goes to infinity, in agreement with the picture of Witten
\cite{Wittheta} for the case of pure gluodynamics.}  
\beq
\label{19g}
S_4 = 
\frac{ 3^3 \cdot 2^7 \cdot \pi^2 p^4}{q^4 N_{f}^5} \, \frac{f_{\pi}^4
E^2}{ M^3} \, \frac{ \left( 1 - \cos \frac{\pi}{2p} \right)^4 }{
\left( 1 - \cos \frac{ 2\pi}{q N_f} \right)^3 } \simeq 
\frac{27}{256} \, \frac{\pi^4 q^2 N_f}{p^4} \, \frac{
f_{\pi}^4  \la \frac{ b \alpha_s }{ 32 \pi} G^2 \ra^2}{ m_{q}^3 
\left| \la \bar{\Psi}\Psi\ra \right|^3 } \; . 
\eeq
Eq.(\ref{19g}) shows that the false vacuum decay is suppressed 
parametrically by a factor $ \sim (\Lambda_{QCD}/m_{q})^3 $,
which should be compared to a factor $ \sim  
(\Lambda_{SYM}/m_{g})^3 $ in the soft SUSY breaking scheme 
\cite{KSS,Shifman}. While the latter ceases to yield a suppression
with approaching the decoupling limit $ m_{g} \gg \Lambda_{SYM} $, 
the former is a real suppression factor for QCD. As should 
be expected, it tends to infinity (i.e. vacua become stable)
when $ m_q $ goes to zero. On the other hand,    
Eq.(\ref{19g}) shows that the parametric suppression 
of the decay is largely overcome due to a numerical 
enhancement. The latter depends crucially on the particular 
values of the integers $ p , q $. In particular, for 
our favorite choice $ p = 11 N_c - 2 N_f , q = 8 $,
Eq.(\ref{19g}) yields a factor $ \simeq 10 $, while for 
$ p = N_c , q = 1 $ (as motivated 
by SUSY, see Sect.3) it is 
approximately two orders of magnitude larger,
but still much smaller than the estimate of \cite{Smilga} for 
the VVW potential. For a discussion of these results, see
\cite{axion}.

To conclude this section, we would like to note that there 
also exist 
other domain walls interpolating between different local 
minima of the effective potential (\ref{8}). The surface 
tension $ \sigma $ and decay rate $ \Gamma $ for these walls
strongly depend on the vacuum states connected by the wall. 

\section{Pseudo-goldstone bosons at different $ \theta $ angles}

In this section we address a few related questions. First, we 
discuss the calculation of the $ \eta' $ mass from the effective 
chiral Lagrangian (\ref{8}) and show that the main contribution
to $ m_{\eta'} $ is given by the conformal anomaly.
We also calculate the $ \theta $ dependence of the $ \eta' $ mass.
Furthermore, it will be shown that for non-zero 
values of $ \theta $, the pseudo-goldstone bosons cease to be 
the pure pseudoscalars, but in addition acquire scalar components.

To study the properties of the pseudo-goldstone 
bosons, we parametrize the chiral matrix (\ref{4}) in the form
\beq
\label{20k}
U = U_{0} \, \exp \left[ i \sqrt{2} \, \frac{\pi^{a} 
\lambda^{a} }{f_{\pi}}  + 
i \frac{ 2}{ \sqrt{N_{f}} } \frac{ \eta'}{ f_{\eta'}}  \right] 
\; ,
\eeq
where $ U_0 $ solves the minimization equations for the effective 
potential (\ref{8}), and the fields $ \pi^{a} \, , \, \eta' $
all have vanishing vacuum expectation values. A simple calculation
of the matrix of second derivatives 
yields the following result for the mass matrix for an arbitrary 
value of $ \theta $:
\bea
\label{20g}
m_{33}^2 &=& \frac{2}{f_{\pi}^2} ( M_u \cos \phi_u + M_d \cos \phi_d )
\nonumber \\
m_{88}^2  &=& \frac{2}{3 f_{\pi}^2} ( M_u \cos \phi_u + 
M_d \cos \phi_d 
+ 4 M_s \cos \phi_s ) \nonumber \\
m_{11}^2 &=& 4 \left( \frac{q}{p} \right)^2 \, 
\frac{E}{f_{\eta'}^2 }
\, N_{f} \, \cos \left( - \frac{q}{p} \theta + \frac{q}{p} \sum 
\phi_{i} + \frac{2 \pi l }{p} \right) + \frac{4}{N_f} 
\frac{1}{f_{\eta'}^2}
\sum_{i= u,d,s} M_{i} \cos \phi_i \nonumber \\
m_{38}^2 &=& m_{83}^2  = \frac{2}{
\sqrt{3}f_{\pi}^2} \left( M_u \cos \phi_u - M_d \cos \phi_d 
\right) \\
m_{31}^2 &=& m_{13}^2 = 
2 \sqrt{ \frac{2}{N_{f}} }\frac{1}{f_{\pi} f_{\eta'}}
\left(  M_u \cos \phi_u - M_d \cos \phi_d \right)  \nonumber \\
m_{81}^2 &=& m_{18}^2  =  2 
\sqrt{ \frac{2}{3 N_{f}} }\frac{1}{f_{\pi} f_{\eta'}}
\left(  M_u \cos \phi_u + M_d \cos \phi_d 
- 2 M_s \cos \phi_s \right)  \nonumber \; , 
\eea
where $ \phi_i $ are solutions of the minimization equation 
stemming from
Eq.(\ref{13}). They depend on $ \theta $ as well as other parameters 
of the effective Lagrangian. 
If the  $ \pi^{0} - \eta - \eta' $ mixing is neglected, $ m_{11} $
coincides with the physical mass of the $ \eta' $. For $ \theta = 0 $
and the particular choice $ p = 3b, q = 8 $, we reproduce in this 
limit the relation
given in \cite{QCD}:
\beq
\label{12}
f_{\eta'}^{2} \, m_{\eta'}^2  = \frac{8}{9 b} N_{f} 
\la \frac{ \alpha_s}{\pi} \, G^2 \ra - \frac{4}{N_{f}} 
\sum_{u,d,s} m_{i} \la \bar{\Psi}_{i} \Psi_{i} \ra + O(m_{q}^{2}) \; . 
\eeq
(The choice $ p = N_c , q = 1 $ would produce a numerically 
close result.)
This mass relation for the $ \eta' $ appears reasonable 
phenomenologically. 
Note that, according to Eq.(\ref{12}), the strange 
quark contributes 30-40 \% of the $ \eta' $ mass. This may lead us 
to expect that chiral corrections 
$ O(m_{s}^2) $ could be quite sizeable. We also note
that in the formal
limit $ N_c \rightarrow \infty, m_q \rightarrow 0 $ 
Eq.(\ref{12}) coincides with
the relation obtained in \cite{2}. In this limit $ m_{\eta'}^2 $ 
scales 
as $ N_{f}/N_{c} $, in agreement with Ref. \cite{WV}. In the 
different 
limit when $ N_c $ goes to infinity at fixed non-zero $ m_q $, 
the result is $  m_{\eta'}^2 = O(m_{q} N_{c}^0 ) $, as for ordinary 
pseudo-goldstone bosons. 
As for the $ \theta $ dependence of 
the $ \eta' $ mass in the same limit, it is given by the third of 
Eq.(\ref{20g}) where the phases $ \phi_i $ implicitly depend on
$ \theta $ through the minimization equation for the effective 
potential (\ref{8}). 

The mass matrix (\ref{20g}) can be used to study the mixing between 
pseudo-goldstone bosons (including also its $ \theta $ 
dependence). Let 
us 
consider the simplest case of the $ \eta - \eta' $ mixing which 
decouples from the $ \pi^0 $ in the isospin limit $ m_u = m_d = m$,
$ \la \bar{d} d \ra = \la \bar{u} u \ra = \la \bar{q} q \ra $. 
For the case $ \theta = 0 $,
it is easy to verify that the mixing matrix 
\bea
\label{300}
m_{\eta-\eta'}^2 = \left( \begin{array}{cl} 
- \frac{4}{3 f_{\pi}^2}( 2 m_s \la \bar{s}
s \ra + m \la \bar{q} q \ra ) &   \frac{4 \sqrt{2}}{ \sqrt{3} f_{\pi} 
f_{\eta'}}
(  m_s \la \bar{s}
s \ra - m \la \bar{q} q \ra ) \\
\frac{4 \sqrt{2}}{ \sqrt{3} f_{\pi} f_{\eta'}}
(  m_s \la \bar{s}
s \ra - m \la \bar{q} q \ra ) &  - \frac{4}{3 f_{\eta'}^2} \sum m_i 
\la \bar{\Psi}_{i} \Psi_{i} \ra + \frac{N_{f} b}{8} \left( \frac{q}{p}
\right)^2 \, \frac{ \la (\alpha_{s}/ \pi) G^2 \ra }{ f_{\eta'}^2 }
\end{array} \right)
\eea
coincide with an accuracy $ O(m_{q}^2) $ with the matrix given
by Veneziano \cite{WV}
\bea
\label{301}
m_{\eta-\eta'}^2 = \left( \begin{array}{cl} 
 \frac{1}{3} ( 4 m_{K}^2 - m_{\pi}^2 ) 
&  -  \frac{2 \sqrt{2}}{ 3 }  (  m_{K}^2 - m_{\pi}^2 ) \\ 
 -  \frac{2 \sqrt{2}}{ 3 }  (  m_{K}^2 - m_{\pi}^2 ) & 
\frac{2}{3} m_{K}^2 + \frac{1}{3} m_{\pi}^2 + \frac{ \chi}{N_c} 
\end{array} \right)
\eea
with the only (but important) difference that the topological
susceptibility $ \sim \chi $ in pure YM theory in the latter 
is substituted by the term proportional to the gluon condensate
in {\it real} QCD in the former. For the particular 
values  $ p = 3b = 11 N_c - 2 N_f , q = 8 $, we may write 
down a QCD analog of the Witten-Veneziano formula:
\beq 
\label{302}
\la \frac{ \alpha_{s}}{ \pi} G^2 \ra = \frac{3b}{8} f_{\eta'}^2 \, 
( m_{\eta'}^2 + m_{\eta}^2 - 2 m_{K}^2 ) + O(m_{q}^2 ) \; .
\eeq
This relation generalizes Eq.(\ref{12}) as it now includes the $ \eta - 
\eta' $ mixing. 

Eqs. (\ref{20g}) can also be used to study other problems related 
to the physics of the pseudo-goldstone bosons. In particular,
we may find the mixing angles in the system $ \pi^0 - \eta - \eta' $
at zero and non-zero angles $ \theta $. Instead of discussing these 
more 
phenomenological issues, we here would like to address another 
interesting aspect of the corresponding physics. Namely, we would 
like to show that the neutral pseudo-goldstone bosons in the 
$ \theta $-vacuum cease to be the pure pseudoscalars, but  instead 
become mixtures of the scalar and pseudoscalar states\footnote{
This fact was previously noted in  the literature
\cite{Huang}.}. To show this, we 
note that the result found for the quark condensates in the $ 
\theta $-vacuum 
\beq
\label{303}
\la \bar{\Psi}_{L i}' \Psi_{R j}' \ra_{\theta}  \equiv 
\la \bar{\Psi}_{L i} \Psi_{R j} \ra_{\theta = 0} \, e^{i \phi_{i} } 
\eeq
can be represented as a chiral rotation of the usual $ \theta = 0 $ 
vacuum:
\beq
\label{304}
\Phi' = U_R \Phi U_{L}^{+} \; \; , \; \; \Phi_{ij} = 
\la \bar{\Psi}_{L i} \Psi_{R j} \ra_{\theta = 0} \; .
\eeq
Under such a rotation, the quark fields transform as 
\beq
\label{305}
\Psi_{R i}' = \left( U_R \right)_{ik} \Psi_{R k} \; \; , \; \; 
\Psi_{Lj}^{+'} = \Psi_{L k}^{+} \left( U^{+} \right)_{kj} \; . 
\eeq
In the ``rotated" basis, the spin content of the pseudo-goldstone
bosons is the standard one. However, the relations (\ref{305})
imply that it will generally 
have a different form in terms of the original unrotated fields.
From Eqs.(\ref{303},\ref{304}) we obtain
\beq
\label{306}
\left( U_{R} U_{L}^{+} \right)_{ii} = e^{i \phi_i} \; \; 
(no \; \; sum \; \; over \; \; i) \; \; \Rightarrow 
\left( U_{R} \right)_{ik} = \delta_{ik} e^{ \frac{i}{2} \phi_i } \; \; , 
\; \; \left( U_{L}^{+} \right)_{kj} = \delta_{kj}     
 e^{ \frac{i}{2} \phi_i } \; .
\eeq
Let us consider e.g. the $ \pi^{0} $ field in the $ \theta $-vacuum.
In the ''rotated" basis, it has the usual spin content. Using the 
correspondence (\ref{305},\ref{306}), we obtain
\beq
\label{307}
| \pi^{0} \ra \sim  | \bar{u}' i \gmf u' - \bar{d}' i \gmf d' \ra = 
\cos \phi_{u} \, | \bar{u} i \gmf u \ra - \sin \phi_u \, | \bar{u} 
u \ra - 
( u \leftrightarrow d )
\eeq
Eq.(\ref{307}) illustrates the phenomenon announced in the 
beginning of this section: In the presence of a non-zero 
angle $ \theta $, the pseudo-goldstone bosons cease to be 
the pure pseudoscalars, but in addition acquire scalar components.
Although in reality $ \theta $ is extremely close to zero, this 
observation is not only of academic interest. The point is that 
in heavy ion collisions one can effectively create, in principle, an 
arbitrary value of $ \theta $ \cite{axion}. In these circumstances,
the scalar admixture in the pseudo-goldstones would be quite
large, and probably could play an important role in dynamics.
    
\section{Further applications and speculations}
\subsection{Axion potential from effective Lagrangian}

One of the interesting implications of the present effective 
Lagrangian
approach 
concerns the possibility to construct a
realistic axion potential 
\cite{axion}
consistent with the known Ward identities of QCD\footnote{
Perhaps, one 
should note that some 
popular {\it Ans\"{a}tze} for the axion potential - such as 
 $V(a)= m_a^2 a^2/2 $ or $V(a)\sim \cos(a/f_a)$ - are at variance 
with the Ward identities of QCD.}.
This may be achieved due to the  
one-to-one correspondence between the 
form of the axion potential
$V(a)$  and the 
vacuum energy $E_{vac}(\theta)$ 
as a function of the  fundamental QCD parameter $\theta$.
Indeed, the axion solution of the strong CP problem suggests
(see e.g. \cite{axrev}) 
that $\theta$ parameter in QCD is promoted
to the  dynamical axion field $\theta \rightarrow a(x)/f_a $, and 
the QCD vacuum energy $E_{vac}(\theta)$ 
 becomes the axion potential $V(a)$. Therefore, 
the problem of analysing $V(a)$ amounts to the 
study of $E_{vac}(\theta)$ in QCD without the axion, 
which is exactly the problem addressed above in the present paper.

Both the local and global properties of the axion potential
can be analysed with this approach. As for the former, we 
note that, as all dimensionful parameters
in our effective Lagrangian are fixed in 
terms of the QCD quark and gluon condensate, the 
temperature dependence of 
the axion mass (and of the entire axion potential) 
can be related with that of
the QCD
condensates whose temperature dependence is understood 
(from lattice or model calculations). 

In particular, the axion mass, which is defined 
as the quadratic coefficient in the expansion of the function
$E_{vac}(\theta)$ at small $\theta$, is proportional to
the chiral condensate: 
$m_a^2(T)\sim m_q \lo\bar{\Psi}\Psi\ro_T/ f_{a}^2$.
Therefore, $m_a^2(T)$ is known as long as $\lo\bar{\Psi}\Psi\ro_T$ 
is known. This statement is exact up to the higher order 
corrections in
$m_q$. We neglect these higher order corrections everywhere for
$T\leq T_c$ ($T_c\simeq 200 \, MeV$ is the critical temperature),
where the chiral condensate is nonzero and gives the most important
contribution to $m_a$. 
For the particular case 
 $N_f=2$
one expects a second order phase transition and, therefore,
$m_a^2\sim\frac{m_q}{f_a^2}\lo\bar{\Psi}\Psi\ro 
\sim |T_c-T|^{\beta}$
for   $T$  near $ T_c \simeq 200 \, MeV $. 
This is exactly where the 
axion mass 
does ``turn on''. The critical exponent in this case 
$\beta\simeq 0.38$,
see e.g. recent reviews \cite{reviews} for a general discussions 
of the 
QCD phase transitions.

The global (topological) structure of the axion potential
appear to be rather complicated, in contrast to what could be 
expected according to simple model potentials such as   
 $V(a)= m_a^2 a^2/2 $ or $V(a)\sim \cos(a/f_a)$.   
In particular, it admits the appearance of additional
local minima of an effective potential.
Thus, the axion
potential may become a
multi-valued function, i.e. there would be two different
values of the axion potential $V_{1,2}(\theta=a/f_a)$
for a fixed $\theta$, which differ by the phase of the chiral
field. For the VWW potential, this happens only at $ \theta \sim
\pi $ for a small isospin breaking, while for the effective 
potential
(\ref{8}) with $ q \neq 1 $ metastable vacua exist for 
any $ \theta $, 
similarly to the case of a softly broken SUSY. 
Interpolating between two minima is the domain wall
that was described in Sect. 5. We stress that it is {\it not}
an axion domain wall, as the value of $ \theta $ does not 
change in this transition. 
Similar domain walls which separate
vacua with different phases of the gluino condensate have been
recently discussed \cite{KSS,KKS} for the SUSY models. 
The appearance of such a domain wall implies an interesting 
dynamics which develops at temperatures {\it below} the chiral
phase transition, which still
has to be explored. 
 
\subsection{Axially disoriented chiral condensate and 
axion search at RHIC}

Another interesting phenomenon amenable to an analysis within 
the effective Lagrangian framework is the possibility 
of production of finite regions of $ \theta \neq 0 $ vacua
in heavy ion collisions \cite{axion}, where the 
chiral fields are misaligned from the true 
vacuum in the axial $ U_{A}(1) $ 
direction. This is somewhat
analogous to the production of the disoriented chiral condensate
(DCC) with a ``wrong" isospin direction (see e.g. \cite{DCC} for 
a review).

Let us briefly recall the reason as to why the DCC could 
be produced and observed in heavy ion collisions. The energy 
density of the DCC is determined by the mass term:
\beq
\label{308}
 E_{\phi}= -\frac{1}{2} Tr( M U + M ^{+}U^{+}) = 
 - 2m  | \la \bar{\Psi}  \Psi  \ra | \cos(\phi)
\eeq
where  we put $m_u=m_d=m$  for simplicity, and $ \phi $ stands for 
the misalignment angle. Thus, the  energy 
difference
between the misaligned state and true vacuum 
with $\phi=0$ is small
and proportional to $m_q$. Therefore, the probability to create 
a state
 with an arbitrary $\phi$ at high temperature 
$ T \sim T_c $ is proportional to
$ \exp[- V( E_{\phi} - E_0)/T]$  and  depends on 
$\phi$ only very weakly, i.e. $ \phi $ is a quasi-flat 
direction. Just after the phase transition when $\la 
\bar{\Psi}  \Psi  \ra $ 
becomes nonzero,
the pion field begins to roll toward $\phi=0$, and of course 
overshoots $\phi=0$.
Thereafter, $\phi$ oscillates. One should expect coherent 
oscillations of the 
$\pi$ meson field which would correspond to a zero-momentum 
condensate of pions.
Eventually, these classical oscillations produce the 
real $\pi$ mesons which
hopefully can be observed at RHIC.

We now wish to generalize this line of reasoning to the 
case when the chiral phases are misaligned in the $ U_{A}(1) $
direction as well. For
arbitrary phases $ \phi_i$  the
energy of a misaligned state differs by a huge 
amount  $\sim E $ from the vacuum energy. 
Therefore, apparently there are no quasi-flat 
misaligned $ U_{A}(1) $ directions 
among $\phi_i$ coordinates, 
which would lead to long wavelength oscillations with production
of a large size domain. However, when
the relevant combination  $(\sum_i\phi_i-\theta)$ from Eq.(\ref{8}) 
is close by an amount $ \sim O(m_q) $ to its vacuum value, a 
Boltzmann suppression due to the term $ \sim E $ is 
absent, and an arbitrary misaligned $|\theta\ra$-state
can be formed. In this case for any $ \theta $ 
the difference in energy
between the true $|\theta\ra$ vacuum
 and a misaligned $|\theta\ra$- state  (when the $\phi_i $
fields are not yet in their final positions 
$\overline{\phi_i(\theta)}$)
 is proportional to 
$m_q$ and very small in close analogy to the DCC case.
 
Once formed, such a domain with $ \theta \neq 0 $
could serve as a source of axions, thus suggesting a 
new possible strategy for the axion search. Below we would like 
to sketch this idea, referring the interested 
reader to Ref. \cite{axion}
for more details.

It is well known that $ \theta $ is a world constant
in the usual infinite volume equilibrium formulation of 
a field theory. The superselection rule ensures that the only 
way to change $ \theta $ under these conditions is to have 
an axion in the theory. However, due to the fact that we do not expect 
to create 
an equilibrium state with an infinite  correlation length 
in heavy ion collisions,
the decay of a $|\theta\ra$-state will also  occur due to the  
Goldstone $U$ fields with specific $CP$-odd correlations\footnote{
A similar phenomenon has been recently discussed in 
Ref.\cite{Pisarski} where the possibility of 
spontaneous parity breaking in QCD at $ T \simeq T_c $ 
was studied using the large $ N_c $
Di Vecchia-Veneziano-Witten  effective chiral Lagrangian.}.
Therefore, two mechanisms of the relaxation of a $|\theta\ra$-state
to the vacuum would compete: the axion one and the 
standard decay
to the Goldstone bosons. In the large volume limit  if 
a reasonably good equilibrium state 
with a large correlation length is created,
the axion mechanism
would win; otherwise, the Goldstone mechanism would win. In any case, 
the result of the decay of a $|\theta\ra$-state
would be very different depending on the presence or 
absence of the axion
field  in Nature\footnote{The possibility
of production of axions in heavy ion collisions
was independently discussed by Melissinos \cite{ion}.}. 
Provided the axion production is strong enough, the axion 
could be detected by using their property of conversion into 
the photons in an external magnetic field \cite{sikivie}.
Thus, heavy ion collisions may provide us with a way to finally 
catch the so far elusive axion. 

\subsection{Early Universe during the QCD epoch}

As was discussed in Sect. 5, the effective Lagrangian approach 
developed in this paper predicts the existence of the 
domain wall excitations in QCD at zero temperature. One may
expect that these domain walls appear also for  non-zero temperatures
$ T< T_c $ where $ T_c $ is the temperature of the chiral phase 
transition. If so, it would be very interesting to 
study this dynamics in the cosmological context. Here we only mention
that the walls discussed above are harmless cosmologically as 
they decay in a proper time \cite{axion}. On the other hand, as 
was noted in \cite{axion}, the dynamics of the decaying domain walls
is an out-of-equilibrium process with $ 100 \% $ violation of $CP$
invariance. This is because the phase of the chiral condensate
in the metastable vacuum is nonzero and of order 1, that  
leads to violation of $CP$ even if $ \theta = 0 $.
(This is not at variance with the Vafa-Witten theorem \cite{VW}
which refers to the lowest energy state only.)    
It was speculated in \cite{axion} that such effects could 
lead to 
a new mechanism for baryogenesis at the QCD scale. Indeed, 
 it appears that
all three famous 
Sakharov criteria \cite{Sakh} could be 
satisfied in the decay of a metastable
state discussed above:\\
1. Such a metastable state is clearly out of thermal
equilibrium; \\
2. CP violation is unsuppressed and proportional
to $  m_um_dm_s \bar{\theta}_{eff}
 ,~\bar{\theta}_{eff} \sim 1$. 
As is known, this is the
most difficult part to satisfy
in the scenario of baryogenesis at the electroweak scale
within the standard model for CP violation; \\
3. The third Sakharov criterion 
is violation of the baryon $(B)$ number.
Of course, the corresponding U(1) is an exact global symmetry
of QCD. However, a ``spontaneous" 
baryon number non-conservation
could arise in this dynamics as  
a result of interactions of fermions with the domain wall.
In this case, baryogenesis
at the QCD scale is feasible. One possible scenario \cite{BHZ}
of such a ``spontaneous" baryogenesis with 
zero net baryon asymmetry is a mechanism based on a charge 
separation, when the anti-baryon charge is concentrated on
the surface of balls of the metastable vacuum 
produced in evolution of domain walls (B-shells).
Rough estimates \cite{BHZ} show that the observed ratio
$ (n_B - n_{\bar{B}})/s \sim 10^{-9} $ can be easily reproduced
in this scenario. Surprisingly, the energy density $ \Omega_{\bar{B}}
$ associated with these B-shells can be close to unity. Therefore,
they can be considered as candidates for dark matter. 
We would like to emphasize that each step in
such a scenario for baryogenesis at the QCD scale could be,
in principle, experimentally tested at RHIC.

\section*{Acknowledgements}

Some parts of this work have been presented at the 
Axion workshop (Gainesville 98) and the conferences
``Continuous advances in QCD" (Minneapolis 98) and ``Lattice-98"
(Boulder). We are 
grateful to the participants of these meetings. In particular,
we have benefited from discussions 
with R. Brandenberger, D. Kharzeev, J. Kim, I. Kogan,
A. Kovner, A. Melissinos, E. Mottola, R. Peccei, 
R. Pisarski, M. Shifman, 
E. Shuryak, P. Sikivie, A. Smilga, M. Stephanov, M. Strassler,
B. Svetitsky, A. Vainshtein, L. Yaffe, V. Zakharov and K. Zarembo.

\clearpage
\appendix
\def\theequation{\thesection.\arabic{equation}}

\section*{Appendix: Fixing $ p/q $ by the ``integrating in"}

\def\thesection{A}
\setcounter{equation}{0}

The purpose of this appendix is to suggest a method which allows
one to fix the number $ p/q $ that appears in the effective 
potential (\ref{6}), provided two plausible assumptions 
are made. One of them is insisting on the standard form of the 
fermion mass term in the effective potential, while the 
other one is the hypothesis of preserving the holomorphic 
properties when a heavy fermion is integrated in/out, see below.  
The approach
developed below closely follows the method of \cite{3} where
a similar problem was addressed for the effective Lagrangian \cite{1} 
for pure YM theory. The essence of this method is to 
consider 
QCD with light fermions as a low energy limit of a theory including 
in addition a heavy fermion, and  
to construct an effective Lagrangian for the latter theory
starting from the effective Lagrangian (\ref{6}). As will be 
shown below, a relation between the 
holomorphic and ``topological"
properties of two Lagrangians is non-trivial, and allows one 
to fix the crucial parameter $ \xi = q/(2p) $ 
entering Eq.(\ref{6}).

The task of constructing such an effective Lagrangian for the theory
with a heavy fermion is  
achieved by using the ``integrating in" technique, developed in the 
context of SUSY theories in Ref.\cite{Int} and reviewed by 
Intriligator and Seiberg \cite{SUSY}.
The integrating in procedure can be viewed as a method of introducing
an auxiliary field into the effective Lagrangian for QCD with 
light flavors.
Using the renormalization group properties of the 
QCD effective Lagrangian in the 
chiral limit $ m_q \rightarrow 0 $, the latter is extended 
to include the auxiliary 
field $ T $, which will be later on identified with the chiral
combination $ \bar{Q}_L Q_R $ of a heavy fermion.

To conform with the notation and terminology of Ref.\cite{Int},
we will call QCD with light quarks and the theory with a heavy fermion
the d-theory (from ``downstairs")
and the u-theory (from
``upstairs"), respectively. The 
effective potential of the d-theory is then
$ W_d + W_{d}^{+} $ with (see Eq.(\ref{7}))
\beq
\label{12i}
W_{d} (h, U)  = \frac{1}{4} \frac{q}{p} h \, Log \left[ \left( 
\frac{h}{c \Lambda_{QCD}^{4}} \right)^{p/q}
\frac{ Det \, U }{ e^{-i\theta} } \right] -  \frac{1}{2}
Tr \, M U \; , 
\eeq
(here $ c $ is a dimensionless numerical coefficient),   
and the summation over all branches 
of the logarithm in the partition function is implied. In this
section Eq.(\ref{12i}) will be understood as representing 
a branch (section) of the multi-valued effective potential,
which corresponds to a lowest energy state for small $ \theta \ll
\pi $. As was shown in \cite{1,3}, this section corresponds to 
the principal branch of the rational function in the  
logarithm in Eq.(\ref{12i}). 
 
We now want to relate \cite{SV,Int} 
the dimensional transmutation parameter
$ \Lambda_{QCD} $ of the d-theory to the scale parameter 
$ \Lambda_{QCD+1} $ of the u-theory including a heavy quark of mass
$ m  \gg \Lambda_{QCD}, \Lambda_{QCD+1}$. 
We assume both parameters to be defined in the $ 
\overline{MS} $ 
scheme, in which no threshold factors arise in 
corresponding matching conditions.
 The matching condition then follows from 
the standard one-loop relations
\bea
\label{lam}
\Lambda_{QCD} &=& M_0 \exp \left( - \frac{8 \pi^2}{ b_{QCD} 
g^{2}(M_{0})}
\right) \; \; , \; \; b_{QCD} \equiv b = \frac{11}{3} \, N_c 
- \frac{2}{3} N_f \; , 
\nonumber \\
\Lambda_{QCD+1} &=& M_0 \exp \left( - \frac{8 \pi^2}{ b_{QCD+1} 
g^{2}(M_{0})}
\right) \; \; , \; \; b_{QCD+1} = 
\frac{11}{3} \, N_c - \frac{2}{3} (N_f +1) \; , 
\eea
and 
the requirement
that the coupling constants of the d- and u- theories coincide
at the decoupling scale $ M_{0} = m $. We obtain
\beq
\label{13i}
\Lambda_{QCD}^4 = \Lambda_{QCD+1}^{4} \left( \frac{m^2}{ 
\Lambda_{QCD+1}^{2}}
\right)^{ 4/(3 b) } \; , 
\eeq
As was explained in Ref.\cite{SV,Int,SUSY}, Eq.(\ref{13i}) reflects
the fact that,
for fixed $ \Lambda_{QCD+1} $, the 
scale parameter $ \Lambda_{QCD} $ characterizes the low energy theory
surviving below the scale $ m $, and thus depends on $ m $.  
In this sense, the constant in the logarithm 
in Eq.(\ref{12i}) also depends on $ m $ 
\beq
\label{14i}
 ( c \Lambda_{QCD}^{4})^{p/q} = 
( c \Lambda_{QCD+1}^{4})^{p/q}  
\, \left( \frac{m}{ \Lambda_{QCD+1} } \right)^{
8 p /(3 b q) } \; .
\eeq
Following Ref.\cite{Int},
we now wish to consider (a particular branch of) the effective 
potential (\ref{12i}) as the result of integrating out the 
auxiliary field $ T $ in the new effective potential 
$ W \equiv W_u - m T $ which corresponds to the u-theory:
$ W_{d}(h,m) = W(h , m , \la T \ra ) $, or
\beq
\label{15i}
W_d = [ W_u - m T ]_{ \la T \ra }  \; ,
\eeq
where $ \la T \ra $ is a solution of the classical equation of 
motion for the auxiliary field $ T $ :
\beq
\label{16i}
\frac{ \partial W_u}{ \partial T } - m = 0 \; .
\eeq
Let us note that, according to Eq.(\ref{15i}), 
$ W_d $ should depend holomorphically 
on $ \la T \ra $. Our assumption is that this is only possible
if an effective potential $ W $ of the u-theory is itself holomorphic 
in the field $ T $.
Furthermore, one can see  
that Eqs. (\ref{15i}),(\ref{16i}) actually define the 
potential $ W_d $ as the Legendre transform of $ W_u $.
Therefore we can find the unknown function $ W_u $ from the 
known potential $ W_d $ by the inverse Legendre transform: 
\beq
\label{17i}
W_u = [ W_d + m T ]_{ \la m \ra } \; , 
\eeq
where $ \la m \ra $ solves the equation
\beq
\label{18i}
\frac{ \partial}{ \partial m } \, ( W_d + m T) = 0 \; .
\eeq
Eq.(\ref{18i}) can be considered as an equation of motion for 
the auxiliary ``field" $ m $. It is important to note 
that Eqs. (\ref{15i} - \ref{18i}) 
imply that $ m $ should be treated as a complex parameter to 
preserve the holomorphic structure of Eq.(\ref{12i}).
When substituted in Eq.(\ref{17i}),
a solution $ \la m \ra $ of Eq.(\ref{18i}) defines the potential
$ W_{u}( h , T , \la m \ra ) $. When this function is found, 
the effective potential $ W $ of the u-theory is defined by the 
relation
\beq 
\label{19i}
W(h , T , m ) = W_{u}(h , T , \la m \ra) - m T \; , 
\eeq
in accord with Eq.(\ref{15i}).

The solution of Eq.(\ref{18i}) is easy to find using Eqs. (\ref{12i}),
(\ref{14i}) :
\beq
\label{20i}
\la m \ra = \frac{2}{3b} \, \frac{h}{T} \;. 
\eeq
Thus, Eq.(\ref{17i}) yields
\beq
\label{21i}
W_u = - \frac{1}{4} \, \frac{q}{p} \, h \log \left[ 
\left( \frac{c \Lambda_{QCD+1}^{4}}{h} \right)^{p/q} \left(
\frac{2}{3b} \, \frac{h}{ \Lambda_{QCD+1} T } \right)^{8p/(3bq)}
\frac{e^{-i \theta}}{ Det \, U} \right]
+ \frac{2}{3b} \, h \; .
\eeq
Finally, Eq.(\ref{19i}) results in the effective potential 
of the u-theory
\beq
\label{22i}
W =  \frac{1}{4} \, \frac{q}{p} \, h \log \left[ \left( \frac{
h}{ c \Lambda_{QCD+1}^{4}} \right)^{ ( 1 - 8/3b)p/q }
\left( \frac{3b}{2} \, \frac{ T}{ c \Lambda_{QCD+1}^3 } 
\right)^{8 p/(3bq)} \frac{ Det \, U}{ e^{-i \theta}}
\right] + \frac{2}{3b} \, h - m T \; .
\eeq
We expect this effective Lagrangian to describe QCD with light quarks 
and the additional
heavy quark, corresponding to the field $ T $, 
such that integrating out $ T $ brings us back to the 
effective Lagrangian (\ref{12i}) for QCD.
Indeed the equation of motion for the field
$ T $ stemming from the effective potential (\ref{22i}) reads
\beq
\label{23i}
m \la T \ra = \frac{2}{3b} \, h \; .
\eeq
Inserting this classical vacuum expectation value (VEV) back to 
Eq.(\ref{22i}) (i.e. integrating out the field $ T $), 
we reproduce the effective potential of the d-theory, Eq.(\ref{12i}).
Note that as Eq.(\ref{23i}) should preserve the $ N_c $ counting
rule, we obtain
$ \la h \ra \sim N_{c}^2 $ , $ b \sim N_c $, 
$ \la T \ra \sim N_c $. The $ N_c $ dependence 
of the VEV $ \la T \ra $ is consistent with the identification
$ \la T \ra \sim \la \bar{Q}_L Q_R \ra $ which will 
be suggested below. 
 
To identify the field $ T $ of the 
effective theory with a corresponding operator of the fundamental 
theory, we note the following. As 
is seen from (\ref{23i}), $ T $ has dimension 3, and thus
should 
describe the VEV of an
operator bilinear in the heavy quark fields. Furthermore, as long as
$ m $ is effectively considered as a complex parameter, this 
operator can only be $ \bar{Q}_L Q_R $ or $ 
\bar{Q}_R Q_L $, in accord with the structure of the mass 
term in the underlying fundamental theory. 
Comparing Eq.(\ref{23i}) with the relation\footnote{
The equation (\ref{25i}) follows from the operator 
product expansions $ \la m \bar{Q} Q \ra = - \la \alpha_s /(12 
\pi) G^2 \ra + O(1/m^2) $ , $  \la m \bar{Q} i \gmf Q 
\ra =  \la \alpha_s /(8  
\pi) G \tilde{G} \ra + O(1/m^2) $.} 
 between the VEV's in the
underlying theory:
\beq
\label{25i}
\la m \bar{Q}_L Q_R  
  \ra =  \frac{ \alpha_s}{24
\pi} \left( - G^2 + i \, \frac{3}{2} G \tilde{G} \right) \; ,
\eeq  
and recalling the definition of the fields $ h \, , \, H $, we 
conclude that 
\beq
\label{26i}
\xi = \frac{4}{3b} \; \; , \; \; \frac{q}{p} = \frac{8}{3b} \; \; 
; \; \; \la T \ra = \la  \bar{Q}_L Q_R \ra \; .  
\eeq
We thus see that the introduction of the heavy quark into the
effective theory fixes the parameter $ \xi $ which enters
the effective Lagrangian (\ref{12i}) for QCD. This has been
obtained by the matching the local holomorphic properties of the 
d- and u-theories within the integrating in procedure. As was 
shown in \cite{3}, the matching of the two theories can also 
be considered at the level of global quantization rules
for the $ h \, , \, \bar{h} $ fields which, for general
values of $ p/q $, are  
different for the theories described by Eqs. (\ref{12i}) and 
(\ref{22i}). This matching agrees with the result 
(\ref{26i}), and provides a self-consistency check of the 
whole procedure of using the integrating in procedure to fix 
the parameter $ q/p = 2 \xi $ which enters the effective potential
(\ref{6}).
Finally, we note that the correspondence 
\beq
\label{27i}
m T \Leftrightarrow m  \bar{Q}_L Q_R \; \; , \; \; 
 \bar{m} \bar{T}  \Leftrightarrow m  \bar{Q}_R Q_L \; .
\eeq
between the operators of the effective and 
underlying theories has the same meaning as Eqs.(\ref{1}), i.e.
the classical field $ T $ describes the VEV of the chiral combination
$ \bar{Q}_L Q_R $ of the full (QCD +1) theory.


\end{document}